\documentclass{revtex4-2}
\usepackage{lineno}
\usepackage[utf8]{inputenc}
\usepackage{amsbsy}
\usepackage{amsopn}
\usepackage{amstext}
\usepackage{amsmath,amsthm,amsfonts,amssymb}
\usepackage[mathcal]{eucal}
\usepackage{mathrsfs}
\usepackage{braket}
\usepackage[english]{babel}
\usepackage{color}
\usepackage{esint}
\usepackage{booktabs}
\usepackage{subfigure}
\usepackage{hologo}
\usepackage{graphicx}
\usepackage{float}
\usepackage{units}
\usepackage{textcomp}
\usepackage{orcidlink}
\DeclareGraphicsExtensions{.png,PNG,.pdf,.PDF}
\usepackage{hyperref}
\usepackage{slashed}
\newcommand{\ie}{\begin{equation}}
\newcommand{\fe}{\end{equation}}
\newcommand{\se}{\begin{eqnarray}}
\newcommand{\ff}{\end{eqnarray}}
\begin{document}
\title{The Dirac oscillator in the curved spacetime of a cloud of strings}
\author{R. R. S. Oliveira\,\orcidlink{0000-0002-6346-0720}}
\email{rubensrso@fisica.ufc.br}
\affiliation{Departamento de F\'isica, Universidade Federal do Cear\'a (UFC), Campus do Pici, C.P. 6030, Fortaleza, CE, 60455-760, Brazil}


\date{\today}

\begin{abstract}
In this paper, we determine the relativistic bound-state solutions for the Dirac oscillator (DO) in the curved spacetime of a cloud of strings in $(3+1)$-dimensions, where such solutions are given by the four-component normalized Dirac spinor and by the relativistic energy spectrum. However, unlike in literature, here, we work with the spacetime in two different forms/configurations, that is, both in its original form and in its modified form. To achieve our objective, we work with the curved DO in spherical coordinates, where we use the tetrad formalism. So, by defining a stationary ansatz for the spinor, we obtain two coupled first-order differential equations, and by substituting one equation into the other, we obtain a second-order differential equation. To analytically and exactly solve this differential equation, we use a change of function and of variable. From this, we obtain the well-known Whittaker equation, whose solution is the Whittaker function. Consequently, we obtain the energy spectrum, which is quantized in terms of the radial quantum number $n$ and the angular quantum number $\kappa$, and explicitly depends on the angular frequency $\omega$ (describes the DO), curvature parameter $a$ (describes the cloud of strings), and on the effective rest mass $m_{\text{eff}}$ (describes the rest mass modified by the curvature of spacetime). Besides, we graphically analyze the behavior of the spectrum as a function of $\omega$ and $a$ for three different values of $n$ and $\kappa$, as well as the behavior of the radial probability density for four different values of $\kappa$, $\omega$, and $a$ (with $n=0$).
\end{abstract}

\maketitle

\section{Introduction}

In 1989, Moshinsky and Szczepaniak introduced into the literature the relativistic version/generalization of the well-known quantum harmonic oscillator (QHO) for spin-1/2 particles (i.e., Dirac fermions), which became known as the Dirac oscillator (DO) \cite{Moshinsky}. In other words, the DO is an exact model of the QHO for spin-1/2 fermions in a high-energy regime (or relativistic regime). In essence, to build the DO, it needs to insert into the free Dirac equation (DE) a nonminimal coupling/substitution, given by: $\Vec{p}\to\Vec{p}-im_0\omega\beta\Vec{r}$, where $\Vec{p}$ is the linear momentum vector/operator, $i=\sqrt{-1}$ is the imaginary unit, $m_0>0$ is the rest mass of the oscillator with angular frequency of oscillation given by $\omega\geq 0$, $\Vec{r}$ is the position vector/operator, and $\beta$ is one of the usual Dirac matrices (therefore, the Hamiltonian of the DO is linear in both $\Vec{p}$ and $\Vec{r}$) \cite{Moshinsky,Bentez,Romero}. Consequently, in the nonrelativistic limit (or low-energy regime), the DO reduces to the QHO (whose Hamiltonian is quadratic in both $\Vec{p}$ and $\Vec{r}$) with a strong spin-orbit coupling (i.e., ``nonrelativistic DO = QHO + spin-orbit coupling'') \cite{Moshinsky}. So, in addition to being an exactly solvable model introduced in the context of many-particle models in relativistic quantum mechanics (RQM), the DO can be interpreted as an interaction of the anomalous magnetic moment (AMM) of a neutral Dirac fermion with an electric field generated by a uniformly charged dielectric sphere \cite{Bentez,Romero}. In 2013, the DO was verified experimentally by Franco-Villafañe et al. \cite{Villafane}, where the experimental setup was based on a microwave system consisting of a chain of coupled dielectric disks. So, since it was introduced into the literature, the DO has been studied extensively in different areas/contexts of physics (in fact, to date, \cite{Moshinsky} has an impressive 967 citations according to Google Scholar). For example, the DO has been studied in thermodynamics \cite{Frassino,Oli5}, physics-mathematics \cite{Bentez,Villalba1,Villalba2}, nuclear physics \cite{Grineviciute}, quantum optics \cite{Bermudez2}, graphene physics \cite{Quimbay,Belouad}, and so on. Besides, the DO has already been studied in the Aharonov-Bohm-Coulomb system \cite{Oli6}, quantum phase transitions \cite{Bermudez3}, nuclear structure calculations \cite{Yang}, spin and pseudo-spin symmetries \cite{deOliveira}, noncommutative geometry \cite{Oli7}, Aharonov-Casher effect \cite{Oli8}, 2D quantum ring \cite{Bakke1}, noninertial effects with cosmic strings \cite{Bakke2,Oli9,Oli10}, Bonnor-Melvin-Lambda spacetime \cite{Oli11}, global monopole spacetime \cite{Bragança}, superstatistical properties \cite{Guvendi}, and so on.

In general relativity (GR), the concept of clouds of strings (or simply cloud of strings or string cloud) refer to a theoretical model in which the energy-momentum content of spacetime is represented by a continuous distribution of one-dimensional strings (classical relativistic strings) instead of point particles (i.e., dust) or perfect fluids, which was first introduced/developed by Letelier in 1979 \cite{Letelier1,Letelier2,Letelier3}. In other words, the (gauge-invariant) string cloud model is a generalization/extension of the dust cloud model; consequently, the source of the gravitational field is now strings and no longer point particles \cite{Letelier1,Letelier2,Letelier3}. According to Letelier \cite{Letelier1,Letelier2,Letelier3}, there are two main reasons to study Einstein's equations coupled with a string cloud (with spherical symmetry). First, the relativistic strings at a classical level can be used to construct good models for many interactions (one type of model is the field theories associated with given action-at-a-distance interactions), and second, the universe can be represented as a collection of extended (nonpoint) objects (so a ``string dust'' cosmology should give us a model to investigate properties related to this fact) \cite{Letelier1,Letelier2,Letelier3}. Furthermore, a quantization of such a model may shed some light on the question of discretization in astronomy \cite{Letelier1,Letelier2,Letelier3}. Still, according to Latelier \cite{Letelier1,Letelier2,Letelier3}, the idea of a cloud of strings as an additional source of the gravitational field allows us to obtain the Schwarzschild solution/metric corresponding to a black hole surrounded by a cloud of spherically symmetric strings.

In other words, although a cloud of strings describes a theoretical model of matter curving spacetime (serving as a versatile toy model for string-like effects in GR) that has not yet been verified/observed (i.e., it is still a theoretical hypothesis), it possesses a spherical metric that is the subject of several interesting theoretical studies in various areas/fields of physics (involving problems/systems with spherical symmetry). That is, since it was introduced into the literature, the cloud of strings has gained a lot of attention and notoriety, such as in quark matter in 5D and $n$-dimensional Kaluza–Klein cosmological model \cite{Yilmaz,Adhav}, quintessence \cite{Costa}, various types/models of gravity (e.g., Lovelock, teleparallel, Rastall, and Einstein–Gauss–Bonnet) \cite{Toledo1,Toledo2,Ditta,Cai,Belhaj}, various types of black hole (e.g., Hayward, magnetic Lovelock, Reissner–Nordström, Kerr, Kerr-Newman, Kerr–Newman–AdS, and charged-rotating-AdS) \cite{Nascimento,Ali,Toledo3,Li,Cao,Toledo4,Sadeghi}, Bardeen solution \cite{Rodrigues}, minimal geometric deformation \cite{Panotopoulos}, shadows and quasinormal modes \cite{Vishvakarma2}, wormholes \cite{Waseem,Gogoi}, generalized Klein–Gordon oscillator (KGO) with a Coulomb-type potential \cite{Saidi}, and so on. Recently, cloud of string have been applied in the study of a black hole coupled with nonlinear electrodynamics \cite{Sudhanshu}, GUP-corrected Schwarzschild black hole \cite{Badawi}, strong gravitational lensing \cite{Vishvakarma2}, thermodynamic and phase transitions of black holes \cite{Singh}, black strings \cite{Deglmann}, Hayward–Letelier-AdS black hole \cite{Fatima}, and so on. Furthermore, it is important to highlight that, unlike the KGO \cite{Saidi} (a relativistic quantum oscillator for spin-0 particles or Klein-Gordon bosons), to date, the DO has not yet been investigated/studied in the curved spacetime generated by a cloud of strings.

The present paper has as its objective to determine the relativistic bound-state solutions for the DO under the gravitational effects of a cloud of strings, i.e., in the $(3+1)$-dimensional curved spacetime of a cloud of strings. In particular, the bound-state solutions are given by the solutions of the curved DO, that is, by the four-component normalized Dirac spinor and by the relativistic energy spectrum (relativistic energy levels or high-energy spectrum). Besides, with the normalized spinor in hand, we also determine another important result (often neglected in the literature), which is the radial probability density (or ``radial probability amplitude''). To achieve our objective, we work with the curved DO in spherical coordinates, where the formalism used to write the DO in a curved spacetime was the tetrad formalism (or spin connection formalism/formulation) of RG (or gravity). Furthermore, and unlike the KGO \cite{Saidi}, here, we work with two different forms/configurations for the spacetime/metric of a cloud of strings, that is, we work with the original form (developed by Letelier \cite{Letelier1,Letelier2,Letelier3}, where the curvature parameter of the cloud of strings modifies both the temporal and radial parts of the line element/metric), as well as with the modified/rescaled form (the curvature parameter modifies only the angular part of the line element/metric \cite{Saidi,Costa}), respectively. That is, Ref. \cite{Saidi} worked with the KGO only with the last form, i.e., the original form was completely ignored (besides, the probability density was also completely ignored, i.e., the scalar or Klein-Gordon wave function was not normalized). Therefore, in order to overcome the limitations of this reference, we propose here a more general study, but for the case of the DO.

The structure of this paper is organized as follows. In Sect. \ref{sec2}, we introduce the DO in the $(3+1)$-dimensional curved spacetime of a cloud of strings in spherical coordinates, where the formalism used to write the DO in a curved spacetime was the tetrad formalism. By defining a stationary ansatz for the spinor, we obtain two coupled first-order differential equations, and by substituting one equation into the other (i.e., decoupling one of the equations), we obtain a second-order differential equation (i.e., a Schrödinger-like equation or the ``decoupled/second-order radial DO''). In Sect. \ref{sec3}, we analytically and exactly solve this differential equation via a change of function and a change of variable. From this, we obtain the well-known Whittaker equation, whose solution is the Whittaker function, as well as a quantization condition. Consequently, we obtain the relativistic energy spectrum for the DO (and anti-DO), where we discuss in detail various important/interesting aspects/characteristics of such a spectrum. Subsequently, we graphically analyze the behavior of this spectrum as a function of the angular frequency $\omega$ and the curvature parameter $a$ for three different values of the radial quantum number $n$ and the angular quantum number $\kappa$. In Sect. \ref{sec4}, we graphically analyze the behavior of the probability density (as a function of the radial distance) for four different values of $\kappa$, $\omega$, and $a$, with $n=0$ (ground state). In Sect. \ref{sec5}, we present our conclusions. Here, we use the system of natural units $(\hslash=c=G=1)$, and the spacetime with signature $(+,-,-,-)$.

\section{The Dirac oscillator in the curved spacetime of a cloud of strings \label{sec2}}

The covariant DO in a $(3+1)$-dimensional generic curved spacetime is written in the following form (in spherical coordinates) \cite{Bakke2,Oli9,Oli10,Oli11,Bragança,Lawrie,Oli12,Oli13,Oli14,Antoine}
\begin{equation}\label{dirac1}
i\gamma^\mu(x)\left(\nabla_\mu(x)+m_0\omega r\gamma^0\delta^r_\mu\right)\psi(r,\theta,\phi,t)=m_0\psi(r,\theta,\phi,t), \ \ (\mu=t,r,\theta,\phi),
\end{equation}
where $\gamma^{\mu}(x)=e^\mu_{\ a}(x)\gamma^a$ are the curved gamma matrices, $e^\mu_{\ a}(x)$ are the tetrads (also called tetrad field, vielbeins or vielbein vectors, whose inverses are given by $e^a_{\ \mu}(x)=(e^\mu_{\ a}(x))^{-1}$), being $\gamma^a=(\gamma^0,\gamma^i)=(\gamma^0,\vec{\gamma})=(\beta,\beta\vec{\alpha})$ the usual/flat gamma matrices (in Cartesian coordinates), with $\beta=\gamma^0$ and $\vec{\alpha}=\beta\vec{\gamma}$ being the usual/original Dirac matrices, $\nabla_\mu (x)=\partial_\mu+\Gamma_\mu(x)$ is the covariant derivative, being $\partial_\mu=\frac{\partial}{\partial x^\mu}=(\partial_t,\partial_r,\partial_\theta,\partial_\phi)=\left(\frac{\partial}{\partial t},\frac{\partial}{\partial r},\frac{\partial}{\partial \theta},\frac{\partial}{\partial \phi}\right)$ the usual partial derivatives (spherical four-gradient), $\Gamma_\mu(x)=-\frac{i}{4}\omega_{ab\mu}(x)\sigma^{ab}$ is the spinorial connection (or spinor affine connection), being $\omega_{ab\mu}(x)$ the spin connection, $\sigma^{ab}=\frac{i}{2}[\gamma^a,\gamma^b]=i\gamma^a \gamma^b$ ($a\neq b$) is a flat antisymmetric tensor (``rank-2 Dirac tensor''), and $\psi$ is the four-component spherical Dirac spinor, respectively. Soon, we will make the connection between $\psi$ and the true Dirac spinor, which can also be called the original Dirac spinor (or Cartesian Dirac spinor), given by $\Psi_D$ or even $\Psi_c$ (something similar, but in polar coordinates and in other curved spacetimes, was done in Refs. \cite{Oli12,Oli13}). Here, we use the Latin indices $(a, b, c, \ldots=0,1,2,3)$ to label the local coordinates system (local reference frame or the Minkowski spacetime) and the Greek indices $(\mu, \nu, \lambda, \ldots=t,r,\theta,\phi)$ to label the general coordinates system (general reference frame or the curved spacetime).

Explicitly, we can rewrite Eq. \eqref{dirac1} as follows (i.e., developing/``opening'' the indexes, as it should be)
\begin{equation}\label{dirac2}
\left[i\gamma^t(x)\partial_t+i\gamma^r(x)(\partial_r+m_0\omega r\gamma^0)+i\gamma^\theta(x)\partial_\theta+i\gamma^\phi(x)\partial_\phi+i\gamma^\mu (x)\Gamma_\mu (x)\right]\psi=m_0\psi,
\end{equation}
where
\begin{equation}\label{contribution1}
\gamma^\mu (x)\Gamma_\mu (x)=\gamma^t(x)\Gamma_t(x)+\gamma^{r}(x)\Gamma_{r}(x)+\gamma^{\theta}(x)\Gamma_{\theta}(x)+\gamma^{\phi}(x)\Gamma_{\phi}(x).
\end{equation}

So, before determining the form of the metric, tetrads, curved gamma matrices, and spin and spinorial connections, we first need to define the line element of our problem. As we discussed in the introduction, here, we will work with the line element of the cloud of strings in its original/true form \cite{Letelier1,Letelier2,Letelier3,Toledo1,Toledo2,Costa,Vishvakarma1,Sudhanshu,Rodrigues} as well as in its modified/rescaled form (i.e., the original metric with the rescaling of the $t$ and $r$ variables) \cite{Saidi,Costa}. In this way, the line element for both cases (only a cloud of strings, i.e., neglecting the Schwarzschild black hole) is given by (in spherical coordinates)
\begin{equation}\label{linelements}
\begin{cases}
ds^2_{original}=(1-a)dt^2-\frac{dr^2}{(1-a)}-r^2(d\theta^2+\sin^2\theta d\phi^2), \\
ds^2_{modified}=dt^2-dr^2-(1-a)r^2(d\theta^2+\sin^2\theta d\phi^2),
\end{cases}
\end{equation}
where $a\geq 0$ (with $a\neq 1$) is a positive integration constant associated with the presence of the string (here, we will call it the curvature parameter and, for convenience, we consider $0\leq a<1$), $-\infty <t<+\infty$ (time coordinate), $0\leq r<+\infty$ (radius or radial distance/coordinate), $0\leq \theta\leq \pi$ (polar angle), and $0\leq\phi<2\pi$ (azimuthal angle). So, from a physical point of view, the (original) line element of the cloud of strings generates a Ricci scalar (or curvature scalar) given by $R=\frac{2a}{r^2}$, that is, the curvature is ``Coulomb potential-like'' and, therefore, is infinite at the origin ($r=0$) and zero/null at infinity ($r\to\infty$) \cite{Letelier1,Letelier2,Letelier3,Toledo1,Toledo2,Costa,Vishvakarma1,Sudhanshu} (i.e., analogous to what happens in the case of global monopolar spacetime \cite{Bragança}). Besides, such curvature depends linearly on $a$, which implies that the larger (smaller) the value of $a$ (but $a\neq 1$), the larger (smaller) the curvature. Clearly, in the absence of the cloud of strings ($a=0$), the curvature is null/zero; consequently, we have only the usual/standard line element in spherical coordinates (in the Minkowski spacetime).

On the other hand (and for convenience), in a ``general'' or compact form (i.e., compacting everything into one), we can rewrite \eqref{linelements}
as follows
 \begin{equation}\label{compactlinelement}
ds^2_s=\alpha^2_s dt^2-\frac{dr^2}{\alpha^2_s}-r^2\beta^2_s(d\theta^2+\sin^2\theta d\phi^2),
\end{equation}
where we define by simplicity two real parameter (i.e., two spacetime parameters) given by $\alpha_s=\sqrt{1-a(1+s)/2}$ ($0<\alpha_s\leq 1$) and $\beta_s=\sqrt{1+a(-1+s)/2}$ ($0<\beta_s\leq 1$), where $s=+1$ ($\alpha_+=\sqrt{1-a}$ and $\beta_+=+1$) is for the original line element ($ds^2_+=ds^2_{original}$), and $s=-1$ ($\alpha_-=+1$ and $\beta_-=\sqrt{1-a}$) is for the modified line element ($ds^2_-=ds^2_{modified}$), respectively.

However, starting from the fact that any line element can be written as $ds^2=g_{\mu\nu}(x)dx^\mu dx^\nu$, where $g_{\mu\nu}(x)$ is the curved metric tensor (curved metric or simply metric), being $g^{\mu\nu}(x)$ its inverse (inverse curved metric), implies that such metrics takes the following form
\begin{equation}\label{metric1}
g_{\mu\nu}(x)=\left(\begin{array}{cccc}
\alpha^2_s & \ 0 & 0 &  0\\ 0 & -\frac{1}{\alpha^2_s} &  0 &  0\\ 0 & \ 0 & -r^2\beta^2_s&  0 \\
0 & \ 0 & 0 & -r^2\beta^2_s\sin^2\theta
\end{array}\right), 
\ \
g^{\mu\nu}(x)=\left(\begin{array}{cccc}
\frac{1}{\alpha^2_s} & \ 0 & 0 &  0\\ 0 & -\alpha^2_s &  0 &  0\\ 0 & \ 0 & -\frac{1}{r^2\beta^2_s}&  0 \\
0 & \ 0 & 0 & -\frac{1}{r^2\beta^2_s\sin^2\theta}
\end{array}\right),
\end{equation}
where they must satisfy the following relations/conditions (since we are using the tetrad formalism) \cite{Bakke2,Oli9,Oli10,Oli11,Lawrie,Oli12,Oli13,Oli14,Antoine}
\begin{eqnarray}
&& g_{\mu\nu}(x)=e^a_{\ \mu}(x)e^b_{\ \nu}(x)\eta_{ab},
\nonumber\\
&& g^{\mu\nu}(x)=e^\mu_{\ a}(x)e^\nu_{\ b}(x)\eta^{ab},
\nonumber\\
&& g^{\mu\sigma}(x)g_{\nu\sigma}(x)=\delta^\mu_{\ \nu}=e^a_{\ \nu}(x)e^\mu_{\ a}(x),
\end{eqnarray}
where $\eta_{ab}=\eta^{ab}=$diag$(+1,-1,-1,-1)$ is the Cartesian Minkowski metric tensor (or flat metric), which must satisfy
\begin{eqnarray}
&& \eta_{ab}=e^\mu_{\ a}(x)e^\nu_{\ b}(x)g_{\mu\nu}(x),
\nonumber\\
&& \eta^{ab}=e^a_{\ \mu}(x)e^b_{\ \nu}(x)g^{\mu\nu}(x),
\nonumber\\
&& \eta^{ac}\eta_{cb}=\delta^a_{\ b}=e^a_{\ \mu}(x)e^\mu_{\ b}(x),
\end{eqnarray}
with $\delta^a_b=\delta^\mu_\nu=$diag$(+1,+1,+1,+1)$ being the $(3+1)$-dimensional Kronecker delta (i.e., the $4\times 4$ identity matrix).

So, starting from the fact that we have a diagonal metric, it is plausible to obtain the tetrads also diagonal. In this way, assuming that $e^\mu_{\ a}(x)=\delta^\mu_{\ a} f^\mu (x)$ (no summation in $\mu$), where $f^\mu (x)$ is a function to be determined (and $\delta^\mu_{\ a}$=diag$(+1,+1,+1,+1)$), we have
\begin{equation}
g^{\mu\nu}(x)=e^\mu_{\ a}(x)e^\nu_{\ b}(x)\eta^{ab}=\delta^\mu_{\ a} f^\mu (x)\delta^\nu_{\ b} f^\nu (x)\eta^{ab}=f^\mu (x)f^\nu (x)\eta^{\mu\nu}, \ \ (\eta^{\mu\nu}=\delta^\mu_{\ a} \delta^\nu_{\ b} \eta^{ab}).
\end{equation}

Now, looking only at the diagonal terms $\mu=\nu$, we have
\begin{equation}
g^{\mu\mu}(x)=[f^\mu (x)]^2\eta^{\mu\mu} \to f^\mu (x)=\sqrt{\vert g^{\mu\mu}(x)\vert},
\end{equation}
where implies $e^\mu_{\ a}(x)=\delta^\mu_{\ a} \sqrt{\vert g^{\mu\mu}(x)\vert}$ (remembering, no summation in $\mu$). Similarly, the inverse tetrads are $e^a_{\ \mu}(x)=\delta^a_{\ \mu} \sqrt{\vert g_{\mu\mu}(x)\vert}$. In fact, these two formulas work well because here there are no terms in the metric off-diagonal (i.e., we have a diagonal metric). In particular, this is in perfect agreement with Ref. \cite{Awad} (as well as with Refs. \cite{Dokuchaev,Villalba1990,Jing}), where was studied various forms of diagonal tetrads that accommodate Black Hole solutions in $f(T)$ gravity with certain symmetries. Therefore, using these two formulas with \eqref{metric1}, the tetrads and their inverses take the form
\begin{equation}\label{tetrads}
e^\mu_{\ a}(x)=\left(\begin{array}{cccc}
\frac{1}{\alpha_s} & \ 0 & 0 &  0\\ 0 & \alpha_s &  0 &  0\\ 0 & \ 0 & \frac{1}{r\beta_s}&  0 \\
0 & \ 0 & 0 & \frac{1}{r\beta_s\sin\theta}
\end{array}\right), 
\ \
e^{a}_{\ \mu}(x)=\left(\begin{array}{cccc}
\alpha_s & \ 0 & 0 &  0\\ 0 & \frac{1}{\alpha_s} &  0 &  0\\ 0 & \ 0 & r\beta_s&  0 \\
0 & \ 0 & 0 & r\beta_s\sin\theta
\end{array}\right).
\end{equation}

Consequently, the curved gamma matrices are given by
\begin{equation}\label{gammamatrices}
\gamma^\mu(x)=\begin{cases}
\gamma^t(x)=e^t_{\ 0}(x)\gamma^0=\frac{1}{\alpha_s}\gamma^0, \\
\gamma^r(x)=e^r_{\ 1}(x)\gamma^1=\alpha_s\gamma^1,  \\
\gamma^\theta(x)=e^\theta_{\ 2}(x)\gamma^2=\frac{1}{r\beta_s}\gamma^2,\\
\gamma^\phi(x)=e^\phi_{\ 3}(x)\gamma^3=\frac{1}{r\beta_s\sin\theta}\gamma^3.
\end{cases}
\end{equation}

Now, we are ready to determine the form of the spin and spinorial connections (i.e., now we have all the ``ingredients'' on hand). So, according to Refs. \cite{Oli9,Oli10,Bragança,Antoine,Oli11,Lawrie,Oli12,Oli13,Oli14}, the spin connection is defined as follows (torsion-free)
\begin{equation}\label{spinconnection}
\omega_{ab\mu}(x)=-\omega_{ba\mu}(x)=\eta_{ac}e^c_{\ \nu}(x)\left[e^\sigma_{\ b}(x)\Gamma^\nu_{\ \mu\sigma}(x)+\partial_\mu e^\nu_{\ b}(x)\right], 
\end{equation}
where $\Gamma^\nu_{\ \mu\sigma}(x)$ are the well-known Christoffel symbols of the second type, whose expression is given by
\begin{equation}\label{Christoffel}
\Gamma^\nu_{\ \mu\sigma}(x)=\Gamma^\nu_{\ \sigma\mu}(x)=\frac{1}{2}g^{\nu\lambda}(x)\left[\partial_\mu g_{\lambda\sigma}(x)+\partial_\sigma g_{\lambda\mu}(x)-\partial_\lambda g_{\mu\sigma}(x)\right].
\end{equation}

Therefore, using the metrics given in \eqref{metric1}, implies that the non-null components of the Christoffel symbols are
\begin{equation}\label{Christoffelsymbols}
\Gamma^\nu_{\ \mu\sigma}(x)= \begin{cases}
\Gamma^r_{\ \theta\theta}(x)=-r\alpha^2_s\beta^2_s,
\\
\Gamma^r_{\ \phi\phi}(x)=-r\alpha^2_s\beta^2_s\sin^2\theta,
\\
\Gamma^\theta_{\ \theta r}(x)=\Gamma^\theta_{\ r\theta}(x)=\frac{1}{r},
\\
\Gamma^\theta_{\ \phi\phi}(x)=-\sin\theta\cos\theta,
\\
\Gamma^\phi_{\ \phi r}(x)=\Gamma^\phi_{\ r\phi}(x)=\frac{1}{r},
\\
\Gamma^\phi_{\ \phi\theta}(x)=\Gamma^\phi_{\ \theta\phi}(x)=\frac{\cos\theta}{\sin\theta}=\cot\theta.
\end{cases}
\end{equation}

Consequently, the non-null components of the spin connection are written as
\begin{equation}\label{spinconnection2}
\omega_{ab \mu}(x)=\begin{cases}
\omega_{12\theta}(x)=-\omega_{21\theta}(x)=-\alpha_s\beta_s,
\\
\omega_{13\phi}(x)=-\omega_{31\phi}(x)=-\alpha_s\beta_s\sin\theta,
\\
\omega_{23\phi}(x)=-\omega_{32\phi}(x)=-\cos\theta,
\end{cases}
\end{equation}
where implies in the following non-null components for the spinorial connection
\begin{equation}\label{spinorialconnection}
\Gamma_\mu(x)=\begin{cases}
\Gamma_\theta(x)=\frac{1}{2}\alpha_s\beta_s\gamma^1\gamma^2, 
\\
\Gamma_\phi(x)=\frac{1}{2}(\alpha_s\beta_s\sin\theta\gamma^1\gamma^3+\cos\theta\gamma^2\gamma^3).
\end{cases}
\end{equation}

Therefore, using the matrices \eqref{gammamatrices} and \eqref{spinorialconnection}, the product \eqref{contribution1} takes the following form
\begin{equation}\label{contributionofthespinorialconnection}
\gamma^\mu (x)\Gamma_\mu (x)=\frac{\alpha_s}{r}\gamma^1+\frac{\cot\theta}{2r\beta_s}\gamma^2.
\end{equation}

In that way, using \eqref{gammamatrices} and \eqref{contributionofthespinorialconnection}, we obtain from \eqref{dirac2} the following equation
\begin{equation}\label{dirac3}
\left\{\frac{i\gamma^0}{\alpha_s}\partial_t+i\alpha_s\gamma^1\left(\partial_r+\frac{1}{r}+m_0\omega r\gamma^0\right)+\frac{i}{r\beta_s}\left[\gamma^2\left(\partial_\theta+\frac{\cot\theta}{2}\right)+\gamma^3\frac{1}{\sin\theta}\partial_\phi\right]-m_0\right\}\psi=0.
\end{equation}

According to Ref. \cite{Villalba1990}, we can simplify Eq. \eqref{dirac3} using a spinor given by (also done in \cite{Dokuchaev})
\begin{equation}\label{psi}
\psi=\frac{\Phi}{r\sqrt{\sin\theta}},
\end{equation}
where $\Phi=\Phi(r,\theta,\phi,t)$ is a ``new'' four-component spinor. Therefore, we have
\begin{equation}\label{dirac4}
\left\{\frac{i\gamma^0}{\alpha_s}\partial_t+i\alpha_s\gamma^1\left(\partial_r+m_0\omega r\gamma^0\right)+\frac{i}{r\beta_s}\left[\gamma^2\partial_\theta+\gamma^3\frac{1}{\sin\theta}\partial_\phi\right]-m_0\right\}\Phi=0.
\end{equation}

So, comparing Eq. \eqref{dirac4} in the absence of the DO and the cloud of strings ($\omega=0$ and $\alpha_s=\beta_s=+1$) with the DE of Ref. \cite{Villalba1990} for $\alpha=0$ and $\xi(r)=r$ (i.e., reducing to Minkowski spacetime in spherical coordinates), we conclude that both equations are practically the same. In fact, the difference is that in Ref. \cite{Villalba1990}, the imaginary unit $i$ does not appear explicitly in the DE, and the mass has a positive sign. We believe this occurred because another representation was used for the flat/usual gamma matrices (in our case, we will use the famous standard or Dirac representation). Thus, it implies that our original Dirac spinor (in \cite{Villalba1990} is given by $\Psi_c$) will be written as $\Psi_D=S\psi$, where $S$ is a unitary operator or unitary/similarity transformation ($S^\dagger S=SS^\dagger=1$), given as follows \cite{Villalba1990}
\begin{equation}\label{S}
S=S(\theta,\phi)=e^{-\frac{\phi}{2}\gamma^1\gamma^2}e^{-\frac{\theta}{2}\gamma^3\gamma^1}U,
\end{equation}
where $U$ is a fixed/constant matrix operator (i.e., it does not depend on $\phi$ and $\theta$) given by
\begin{equation}\label{U}
U=\frac{1}{2}[\gamma^1\gamma^2+\gamma^3\gamma^1+\gamma^2\gamma^3+1].
\end{equation}

According to \cite{Villalba1990}, $S$ can be written as: $S=\frac{1}{2}$diag$(Z,Z)$, where $Z=e^{-\frac{i\phi}{2}\sigma^3}e^{-\frac{i\theta}{2}\sigma^2}[1+i(\sigma^1+\sigma^2+\sigma^3)]$ (see page 719). In fact, we can easily achieve this by using an important/fundamental relation that Pauli matrices must satisfy, given by $\sigma^i\sigma^j=\delta^{ij}+i\epsilon^{ijk}\sigma^k$, or better, $\sigma^i\sigma^j=\delta^{ij}+i\epsilon^{ij}_k\sigma^k$  (this is valid regardless of the representation of the gamma matrices) \cite{Moshinsky,Greiner,Strange}. So, knowing that the gamma matrices of \cite{Villalba1990} are given by $\gamma^i$=off-diag$(\sigma^i,\sigma^i)$, implies that the products of the gamma matrices are written as: $\gamma^1\gamma^2=i\Sigma^3$, $\gamma^3\gamma^1=i\Sigma^2$, and $\gamma^2\gamma^3=i\Sigma^1$, where $\Sigma^3$=diag$(\sigma^3,\sigma^3)$, $\Sigma^2$=diag$(\sigma^2,\sigma^2)$, and $\Sigma^1$=diag$(\sigma^1,\sigma^1)$, i.e., $\vec{\Sigma}=(\Sigma^1,\Sigma^2,\Sigma^3)$=diag$(\vec{\sigma},\vec{\sigma})$, being $\sigma^i$ ($i=1,2,3$) or $\vec{\sigma}=(\sigma^1,\sigma^2,\sigma^3)$ the well-known $2\times 2$ Pauli matrices \cite{Greiner,Strange}. However, as we will use here the gamma matrices in the standard/Dirac representation, given by $\gamma^i$=off-diag$(\sigma^i,-\sigma^i)$ \cite{Greiner,Strange}, we have $\gamma^1\gamma^2=-i\Sigma^3$, $\gamma^3\gamma^1=-i\Sigma^2$, and $\gamma^2\gamma^3=-i\Sigma^1$ (in particular, this can be generated from the spatial components of $\sigma^{ab}$, that is, $\sigma^{ij}=i\gamma^i \gamma^j=\epsilon^{ij}_k \Sigma^k=2\epsilon^{ij}_k S^k$ ($=2\times$Levi-Civita symbol$\times$spin vector/operator), where $i,j,k=1,2,3$, with $i$, $j$, $k$ all distinct. Consequently, we can called $\sigma^{ab}$ the ``spin tensor''). Therefore, in our case, $S$ must be written with the opposite signs, that is (in particular, the two exponentials are in full agreement with \cite{Schluter,Schluter2})
\begin{equation}\label{S2}
S=e^{\frac{\phi}{2}\gamma^1\gamma^2}e^{\frac{\theta}{2}\gamma^3\gamma^1}U=e^{-\frac{i\phi}{2}\Sigma^3}e^{-\frac{i\theta}{2}\Sigma^2}U,
\end{equation}
where
\begin{equation}\label{U2}
U=\frac{1}{2}[-\gamma^1\gamma^2-\gamma^3\gamma^1-\gamma^2\gamma^3+1]=\frac{1}{2}[i\Sigma^3+i\Sigma^2+i\Sigma^1+1]=\frac{1}{2}[1+i(\Sigma^1+\Sigma^2+\Sigma^3)].
\end{equation}

Besides, it is important to highlight that one of the authors of Ref. \cite{Villalba1990} (i.e., Villalba) has another paper (published 4 years later \cite{Villalba1994}) in which he worked with the DE in a curved spacetime, where a unitary operator also appears, given by $S$ (``unfortunately'', the same letter $S$ was used to symbolize the unitary operator). However, this $S$ is different from the one in Ref. \cite{Villalba1990} by a factor given by $\frac{1}{r\sqrt{\sin\theta}}$. That is, in \cite{Villalba1990}, $S$ does not depend on $\frac{1}{r\sqrt{\sin\theta}}$ (it is the spherical spinor that depends), while in \cite{Villalba1994}, $S$ does depend (but the spherical spinor does not). In other words, these two papers used equivalent methods to work with the DE in a curved spacetime in spherical coordinates (however, the final form of the original Dirac spinor $\Psi_c$ is basically the same for both cases/methods, as it should be). On the other hand, investigating a little deeper the origin of the operator $S$ (such as shown in \cite{Villalba1994}, but without depending on $U$), everything indicates that Refs. \cite{Schluter,Schluter2} was the first to introduce it in the literature in 1983, that is, long before Refs. \cite{Villalba1990,Villalba1994}. So, comparing Eq. \eqref{dirac4} in the absence of the DO and the cloud of strings ($\omega=0$ and $\alpha_s=\beta_s=+1$) with the DE of Ref. \cite{Villalba1994} for $V(r)=0$ (i.e., absence of the Coulomb potential), we conclude that both equations are practically the same. Besides, according to Refs. \cite{Villalba1990,Villalba1994,Schluter}, the periodicity conditions for the spinors $\Psi_D$ and $\Phi$ (or $\psi$, since $\psi=\frac{\Phi}{r\sqrt{\sin\theta}}$) are $\Psi_D(\phi+2\pi)=\Psi_D(\phi)$ and $\Phi(\phi+2\pi)=-\Phi(\phi)$. It is also important to mention that in polar coordinates (in the Minkowski or curved spacetime), the operator $S$ is simply given by $S(\phi)=e^{-\frac{i\phi\sigma^3}{2}}$ \cite{Villalba2,Oli10,Oli11,Oli12,Oli13,Oli14,VillalbaandPino} or even $S(\phi,\rho)=\frac{1}{\sqrt{\rho}}e^{-\frac{i\phi\sigma^3}{2}}$ \cite{Villalba1,Schluter}.

Now, our next step is to perform a separation of variables in the DE \eqref{dirac4}. With this, we hope to obtain a purely radial equation (radial differential equation), that is, one equation that depends only on $r$. Next, we will transform this equation into a second-order differential equation, and then, we will try to obtain its solutions (i.e., the bound-state solutions). To achieve this objective, we can use an approach made in Ref. \cite{Dokuchaev} (also made in \cite{Villalba1990,Villalba1994,Schluter}), which is to define a ``$K$-operator'' from the angular part of the equation (everything indicates that Schrödinger were the first (or one of the first) to study this operator through DE in a central field \cite{Schrodinger,Brill}). So, defining the following $K$-operator \cite{Dokuchaev,Schluter}
\begin{equation}
K=K(\theta,\phi)=\gamma^0\gamma^1\left[\gamma^2\partial_\theta+\gamma^3\frac{1}{\sin\theta}\partial_\phi\right],
\end{equation}
Eq.\eqref{dirac4} will be rewritten as
\begin{equation}\label{dirac5}
\left\{\frac{i\gamma^0}{\alpha_s}\partial_t+i\alpha_s\gamma^1\left(\partial_r+m_0\omega r\gamma^0\right)+\frac{i}{r\beta_s}\gamma^0\gamma^1 K-m_0\right\}\Phi=0,
\end{equation}
or yet
\begin{equation}\label{dirac6}
\left\{\frac{i\gamma^0}{\alpha_s}\partial_t+i\alpha_s\gamma^1\left(\partial_r+m_0\omega r\gamma^0-\frac{1}{r\alpha_s\beta_s}\gamma^0 K\right)-m_0\right\}\Phi=0,
\end{equation}
or better
\begin{equation}\label{dirac7}
\left\{\frac{i\gamma^0}{\alpha_s}\partial_t+i\alpha_s(\vec{\gamma}\cdot\vec{e}_r)\left(\partial_r+m_0\omega r\gamma^0-\frac{1}{r\alpha_s\beta_s}\gamma^0 K\right)-m_0\right\}\Phi=0,
\end{equation}
where we use the fact that $\gamma^0\gamma^0\gamma^1\gamma^1=-\gamma^0\gamma^1\gamma^0\gamma^1=-(\gamma^0)^2(\gamma^1)^2=-(+1)(-1)=+1$ and $\gamma^0\gamma^1=-\gamma^1\gamma^0$, and $\gamma^1=\gamma^r=(\vec{\gamma}\cdot\vec{e}_r)$ \cite{Greiner,Strange}, where $\vec{e}_r$ is the radial unit vector (it will soon become clear why we use this). So, according to \cite{Dokuchaev}, the $K$-operator must satisfy $K\Phi=k\Phi$, where the real parameter $k=0,\pm 1,\pm 2,\ldots$ describes the eigenvalues of $K$ (however, according to \cite{Schluter}, is given by $k=k_\pm=\pm 1,\pm 2,\pm 3,\ldots$. We will soon see that there is no zero value for $k$, that is, should really be $k=\pm 1,\pm 2,\pm 3,\ldots$). Therefore, using this relation in \eqref{dirac7}, we have
\begin{equation}\label{dirac8}
\left\{\frac{i\gamma^0}{\alpha_s}\partial_t+i\alpha_s(\vec{\gamma}\cdot\vec{e}_r)\left(\partial_r+m_0\omega r\gamma^0-\frac{k}{r\alpha_s\beta_s}\gamma^0\right)-m_0\right\}\Phi=0.
\end{equation}

Now, we need to define the form of the spinor $\Phi$  (i.e., a stationary ansatz or a ``standard spinor'' for a system described by DE or DO in spherical coordinates). Here, we will not use the spinor from Ref. \cite{Dokuchaev} because it is ``ill-defined'' (i.e., the angular part of the spinor of this paper is equal for the upper and lower components, which cannot be. Besides, we will see that the value $k=0$ is also not allowed). So, according to Ref. \cite{Strange} (perhaps the only and best textbook that studies the spherical DO with surprising details), we can define the spinor $\Phi$ through the following ansatz (see pages 103, 174, 229, and 269 of this reference, and page 2013 of Ref. \cite{Schluter})
\begin{equation}\label{spinor}
\Phi(r,\theta,\phi,t)=e^{-iEt}\left(
           \begin{array}{c}
           u(r)\chi^{m_j}_{\kappa}(\theta,\phi)\\
            iv(r)\chi^{m_j}_{-\kappa}(\theta,\phi)\\
           \end{array}
         \right),
\end{equation}
where $E$ ($=E^\pm=\pm\vert E^\pm\vert$) is the relativistic total energy (or relativistic energy spectrum) of the system (or stationary bound states of the DO), $u(r)$ and $v(r)$ are real radial functions or ``upper and lower radial components'' (we do not use $g(r)$ and $f(r)$ because in our case there is already $r$ in the denominator, which appears in \eqref{psi}), $\chi^{m_j}_{\pm\kappa}(\theta,\phi)$ are the spin-angular functions (and are two-component spinors), $m_j=\pm\frac{1}{2},\pm\frac{3}{2},\ldots$ are the eigenvalues (called of total angular momentum projection quantum number or even total magnetic quantum number) of $J_z$ ($z$-component/projection of total angular momentum $\vec{J}$), and obeys the relation $J_z\Phi=m_j\Phi$, and $\kappa$ are the eigenvalues (we can call it simply the angular quantum number) of the operator $-K$, or better, $K\Phi=-\kappa\Phi$ (this is also in accordance with \cite{Schluter}), where $-\kappa$ are the eigenvalues of $K$ (we will soon see that it is the $K$-operator itself), respectively. With respect to the operator $K$, is written as $K=\gamma^0(\vec{\Sigma}\cdot\vec{L}+1)=(\vec
J^2-\vec{L}^2-\vec{S}^2+1)$, or better, $\vec{K}$=diag$(+(\vec{L}\cdot\vec{\sigma}+1),-(\vec{L}\cdot\vec{\sigma}+1))$, where $\vec{J}=\vec{L}+\vec{S}$ is the total angular momentum (for the whole universe, is conserved), being $\vec{L}$ and $\vec{S}$ (=$\vec{\Sigma}/2$) the orbital and spin angular momenta (i.e., $\vec{J}$ is the vector sum of orbital and spin angular momenta) \cite{Strange}. Now, with respect to the parameter $\kappa$, is given by $\kappa=\kappa_\mp=\mp(j+1/2)=\mp 1,\mp 2,\ldots$ (i.e., $\kappa$ can only be negative or positive), where $j=l\pm 1/2$ ($j$ is called the total angular momentum quantum number and $l$ the azimuthal quantum number or orbital quantum number), being that $j$ and $l$ arise from the following relations $\vec{J}^2\Phi=j(j+1)\Phi$ and $\vec{L}^2\Phi=l(l+1)\Phi$ \cite{Strange}. That is, for $j=l-1/2$ ($l=1,2,3,\ldots$), we have $\kappa=l>0$ (i.e., $\kappa$ is the positive orbital angular momentum), and for $j=l+1/2$ ($l=0,1,2,\ldots$), we have $\kappa=-l-1<0$ (i.e., $\kappa$ is the negative or null orbital angular momentum) \cite{Strange}.

So, using \eqref{spinor} and knowing that in the standard/Dirac representation, the zero and vector gamma matrices $\gamma^0$ and $\vec{\gamma}$ (time and spatial components of $\gamma^a$) are written as follows \cite{Greiner,Strange}
\begin{equation}
\gamma^0=\left(
    \begin{array}{cc}
      1\ & 0 \\
      0\ & -1 \\
    \end{array}
  \right), \ \  \vec{\gamma}=\left(
    \begin{array}{cc}
      0 & \vec{\sigma} \\
      -\vec{\sigma} & \ 0 \\
    \end{array}
  \right),
\end{equation}
we obtain from \eqref{dirac8} two coupled first-order differential equations, given by
\begin{eqnarray}
&& \left(\frac{E}{\alpha_s}-m_0\right)u(r)\chi^{m_j}_{\kappa} (\theta,\phi)+\alpha_s i(\vec{\sigma}\cdot\vec{e}_r)\left(\frac{d}{dr}-m_0\omega r+\frac{k}{r\alpha_s\beta_s}\right)iv(r)\chi^{m_j}_{-\kappa} (\theta,\phi)=0,
\label{EDO1}\\
&& -\alpha_s i(\vec{\sigma}\cdot\vec{e}_r)\left(\frac{d}{dr}+m_0\omega r-\frac{k}{r\alpha_s\beta_s}\right)u(r)\chi^{m_j}_{\kappa} (\theta,\phi)-\left(\frac{E}{\alpha_s}+m_0\right)iv(r)\chi^{m_j}_{-\kappa} (\theta,\phi)=0,
\label{EDO2}
\end{eqnarray}
or better
\begin{eqnarray}
&& \left(\frac{E}{\alpha_s}-m_0\right)u(r)\chi^{m_j}_{\kappa} (\theta,\phi)-\alpha_s (\vec{\sigma}\cdot\vec{e}_r)\chi^{m_j}_{-\kappa} (\theta,\phi)\left(\frac{d}{dr}-m_0\omega r+\frac{k}{r\alpha_s\beta_s}\right)v(r)=0,
\label{EDO3}\\
&& \left(\frac{E}{\alpha_s}+m_0\right)v(r)\chi^{m_j}_{-\kappa} (\theta,\phi)+\alpha_s (\vec{\sigma}\cdot\vec{e}_r)\chi^{m_j}_{\kappa} (\theta,\phi)\left(\frac{d}{dr}+m_0\omega r-\frac{k}{r\alpha_s\beta_s}\right)u(r)=0.
\label{EDO4}
\end{eqnarray}

Still according to Ref. \cite{Strange} (page 59), the matrix $(\vec{\sigma}\cdot\vec{e}_r)=\sigma_r$ (with $\sigma_r^2=1$) must satisfy the following relations: $\sigma_r\chi^{m_j}_{\kappa} (\theta,\phi)=-\chi^{m_j}_{-\kappa} (\theta,\phi)$ and $\sigma_r\chi^{m_j}_{-\kappa} (\theta,\phi)=-\chi^{m_j}_{\kappa} (\theta,\phi)$ (or simply $\sigma_r\chi^{m_j}_{\pm\kappa} (\theta,\phi)=-\chi^{m_j}_{\mp\kappa} (\theta,\phi)$). Therefore, using these relations in \eqref{EDO3} and \eqref{EDO4}, we can now obtain two first-order differential equations coupled only by the radial components of the spinor, given by
\begin{eqnarray}
&& \left(E-\frac{m_0}{\alpha_s}\right)u(r)+\left(\frac{d}{dr}-m_0\omega r+\frac{k}{r\alpha_s\beta_s}\right)v(r)=0,
\label{EDO5}\\
&& \left(E+\frac{m_0}{\alpha_s}\right)v(r)-\left(\frac{d}{dr}+m_0\omega r-\frac{k}{r\alpha_s\beta_s}\right)u(r)=0,
\label{EDO6}
\end{eqnarray}
where we must have $E\neq \pm m_0/\alpha_s$. Otherwise, we will not have bound-state solutions (that is, we will not have a quantized energy spectrum, as well as a normalized spinor).

So, in the absence of the cloud of strings ($\alpha_s=\beta_s=+1$) and doing $k\to -\kappa$ (with $E=W$), we verified that Eqs. \eqref{EDO5} and \eqref{EDO6} are exactly the same as in Ref. \cite{Strange} (see page 271), i.e., our general case (due to the cloud of strings) reduces to the usual/particular case (as it should be) since we take $k=-\kappa$ ($=\pm 1,\pm 2,\ldots$). Therefore, we conclude that the parameter $k$ is actually $-\kappa$. That is, the $K$-operator of Ref. \cite{Dokuchaev} (as well as of \cite{Schluter}) is actually the operator $K$ of Ref. \cite{Strange} (or better, written in another way, since the DO started from a symmetrically spherical curved spacetime where the original Dirac spinor depends on the unitary/similarity transformation matrix/operator $S$). Consequently, any integer value of $k$ is permissible except $k=0$, i.e., the eigenvalues of $K$ can take on all integral values except zero \cite{Strange}. Several papers that also agree with this (i.e., $k$ cannot have the value zero, and is written as $k=\pm(j+1/2)$), we can cite Refs. \cite{Villalba1990,Villalba1994,Schluter,Schrodinger,Brill,Unruh,Soffel,Audretsch,Pekeris,Belgiorno,Smoller,Abedi,Neznamov}. Besides, for $\omega=0$ and $\alpha_s=\beta_s=+1$ with $E\to E-V(r)$ and $k\to-\kappa$, we obtain the DE with a spherical potential (see page 230 of Ref. \cite{Strange}, and page 2013 of Ref. \cite{Schluter}). Therefore, using $k=-\kappa$, we have
\begin{eqnarray}
&& \left(E-\frac{m_0}{\alpha_s}\right)u(r)+\left(\frac{d}{dr}-m_0\omega r-\frac{\kappa}{r\alpha_s\beta_s}\right)v(r)=0,
\label{EDO7}\\
&& \left(E+\frac{m_0}{\alpha_s}\right)v(r)-\left(\frac{d}{dr}+m_0\omega r+\frac{\kappa}{r\alpha_s\beta_s}\right)u(r)=0.
\label{EDO8}
\end{eqnarray}

Now, we need to obtain/build a second-order differential equation from Eqs. \eqref{EDO7} and \eqref{EDO8}. In particular, this is done by decoupling one of the equations or even both equations (however, here, such as done in \cite{Strange}, just one is enough). Therefore, isolating $v(r)$ in \eqref{EDO8} and substituting it into \eqref{EDO7}, we obtain the following second-order differential equation (i.e., a Schrödinger-like equation or the ``decoupled/second-order radial DO'') for $u(r)$
\begin{equation}\label{EDO9}
\left[\frac{d^2}{dr^2}-\frac{\kappa(\kappa+\alpha_s\beta_s)}{r^2\alpha_s^2\beta_s^2}-(m_0 \omega r)^2+E_\kappa\right]u(r)=0,
\end{equation}
where we define
\begin{equation}\label{energy}
E_\kappa=\left(E^2-\frac{m^2_0}{\alpha_s^2}-m_0\omega\frac{2\kappa-\alpha_s\beta_s}{\alpha_s\beta_s}\right)>0.
\end{equation}

In particular, in the absence of the cloud of strings ($\alpha_s=\beta_s=+1$), we recover the usual second-order differential equation of the DO in the $(3+1)$-dimensional Minkowski spacetime (as it should be) \cite{Strange} (see page 271). However, unlike \cite{Strange}, where $\kappa(\kappa+1)=l(l+1)$, here, it is not possible to use such a relation since we have the presence of the product $\alpha_s\beta_s$ (that is, it would only be possible if $\alpha_s=\beta_s=+1$ or even $\alpha_s\beta_s=+1$). Indeed, for $\kappa=l>0$, we have $l(l+\alpha_s\beta_s)$, but for $\kappa=-l-1<0$, we have $(-l-1)(-l-1+\alpha_s\beta_s)=(l+1)(l+1-\alpha_s\beta_s)\neq l(l+\alpha_s\beta_s)$. In this way, here, we have to solve Eq. \eqref{EDO9} individually for positive and negative $\kappa$, respectively. However, we can avoid this (``problem'') if we change the form of Eq. \eqref{EDO9}, as Ref. \cite{Bragança} did for the case of the DO in the global monopole spacetime (we will do this shortly). Furthermore, it is important to highlight that if we had considered $a>1$, we would have complex values for $\alpha_s$ and $\beta_s$ (since $\alpha_+=\sqrt{1-a}$ and $\beta_-=\sqrt{1-a}$) and, consequently, we would have a complex second-order differential equation. However, since here we want to have a real energy spectrum (i.e., a physically consistent spectrum with the stationary bound states of the DO), we should only consider $a<1$.

\section{Relativistic Bound-state solutions: Dirac spinor and the relativistic energy spectrum \label{sec3}}

In this section, we will obtain the relativistic bound-state solutions for the DO, which are given by the Dirac spinor and the relativistic energy spectrum. To achieve this objective, we need to solve Eq. \eqref{EDO10}. So, according to Ref. \cite{Bragança}, we can change the form of Eq. \eqref{EDO9} for a more convenient way by writing $u(r)$ as follows
\begin{equation}\label{u1}
u(r)=\frac{f(r)}{\sqrt{r}}.
\end{equation}

Therefore, by making a change of function through $\eqref{u1}$, Eq. \eqref{EDO9} will be rewritten in the following form
\begin{equation}\label{EDO10}
\left[\frac{d^2}{dr^2}-\frac{1}{r}\frac{d}{dr}+\frac{3}{4r^2}-\frac{\kappa(\kappa+\alpha_s\beta_s)}{r^2\alpha_s^2\beta_s^2}-(m_0 \omega r)^2+E_\kappa\right]f(r)=0.
\end{equation}

Now, doing a change of variable in \eqref{EDO10} through a variable given by $y=m_0\omega r^2$ \cite{Bragança,Strange}, we have
\begin{equation}\label{EDO11}
\frac{d^2f(y)}{dy^2}+\frac{(1-4\mu^2)}{4y^2}f(y)+\frac{\nu}{y}f(y)-\frac{1}{4}f(y)=0,
\end{equation}
where we define (real) new parameters given by
\begin{equation}\label{newparameters}
\mu=\frac{\sqrt{\alpha^2_s\beta^2_s+4\kappa(\kappa+\alpha_s\beta_s)}}{4\alpha_s\beta_s}>0, \ \ \nu=\frac{E_\kappa}{4m_0\omega}>0.
\end{equation}

According to Refs. \cite{Bragança,Arfken}, Eq. \eqref{EDO11} is the well-known Whittaker equation (modeled by real parameters $\nu$ and $\mu$) and $f(y)$ is the Whittaker function $M_{\nu,\mu}(y)$, which can be written in terms of the confluent hypergeometric function of the first kind $_{1}F_{1}(y)$ in the following form
\begin{equation}\label{Whittaker}
f(y)=M_{\nu,\mu}(y)=Cy^{\mu+1/2}{e^{-y/2}}_{1}F_{1}\left(\mu-\nu+\frac{1}{2},2\mu+1;y\right),
\end{equation}
where $C$ is a normalization constant. So, using \eqref{Whittaker} and $y=m_0\omega r^2$, we can write the function \eqref{u1} in terms of $y$ as
\begin{equation}\label{u2}
u(y)=C(m_0\omega)^{1/4}y^{\mu+1/4}{e^{-y/2}}_{1}F_{1}\left(\mu-\nu+\frac{1}{2},2\mu+1;y\right).
\end{equation}

In particular, in the absence of the cloud of strings ($\alpha_s=\beta_s=+1$), where $\mu=l/2+1/4$, and defining a ``new'' normalization constant given by $A=(m_0\omega)^{1/4}C$, we obtain exactly (as it should be) the radial function for the case of the usual DO from Ref. \cite{Strange} (see page 271 - (9.32a)). However, returning the original coordinate $r$ (with $y=m_0\omega r^2$), \eqref{u1} takes the form
\begin{equation}\label{u3}
u(r)=C(m_0\omega)^{1/4}(\sqrt{m_0\omega}r)^{2\mu+1/2}{e^{-m_0\omega r^2/2}}_{1}F_{1}\left(\mu-\nu+\frac{1}{2},2\mu+1;m_0\omega r^2\right).
\end{equation}

Now, with respect to the energy spectrum, it can be obtained from a relation (or quantization condition) given by $\mu-\nu+1/2=-n$ ($n=0,1,2,\ldots$) \cite{Bragança,Strange}, that is, the first term of the argument of the confluent hypergeometric function must be equal to a negative integer, which is a mandatory requirement for us to have a finite series or a polynomial of degree $n$ (with this, the spinor can then be normalized later). Consequently, this implies that $\nu$ can be written/defined as $\nu=\nu_n=n+\mu+1/2$, i.e., the parameter $\nu$ is quantized since it depends on $n$ (of course, also on $\kappa$). Therefore, using this relation (also of quantization) with $\nu$ and $\mu$ given in \eqref{newparameters}, as well as $E_\kappa$ given in \eqref{energy}, we then obtain the following relativistic energy spectrum (or quantized energy spectrum) for the DO in the curved spacetime of a cloud of strings
\begin{equation}\label{spectrum}
E=E^{\pm}_{n,\kappa}=\pm\sqrt{\frac{m^2_0}{\alpha^2_s}+4m_0\omega\left[n+\frac{\sqrt{\alpha^2_s\beta^2_s+4\kappa(\kappa+\alpha_s\beta_s)}+2\kappa}{4\alpha_s\beta_s}+\frac{1}{4}\right]}=\pm \vert E_{n,\kappa}\vert=\pm \vert E_{n,\kappa}^\pm\vert,
\end{equation}
or in terms of the original and modified spectrum, i.e., the spectrum associated with the original and modified metric, such as (remembering that $\alpha_+=\sqrt{1-a}$ and $\beta_+=+1$ is for the original metric and $\alpha_-=+1$ and $\beta_-=\sqrt{1-a}$ is for the modified metric)
\begin{eqnarray}
&& E^{original}_{n,\kappa}=E^{s=+1}_{n,\kappa}=\pm\sqrt{\frac{m^2_0}{\alpha^2_+}+4m_0\omega\left[n+\frac{\sqrt{\alpha^2_++4\kappa(\kappa+\alpha_+)}+2\kappa}{4\alpha_+}+\frac{1}{4}\right]}, \ \ (\kappa=\mp (j+1/2)),
\label{spectrum2}\\
&& E^{modified}_{n,\kappa}=E^{s=-1}_{n,\kappa}=\pm\sqrt{m^2_0+4m_0\omega\left[n+\frac{\sqrt{\beta^2_-+4\kappa(\kappa+\beta_-)}+2\kappa}{4\beta_-}+\frac{1}{4}\right]}, \ \ (\kappa=\mp (j+1/2)),
\label{spectrum3}
\end{eqnarray}
where $E>0$ ($E^+_{n,\kappa}=+\vert E_{n,\kappa}\vert>0$) describes/represents the positive-energy states/solutions, and $E<0$ ($E^-_{n,\kappa}=-\vert E_{n,\kappa}\vert<0$) the negative-energy states/solutions, respectively \cite{Strange,Greiner,Thomson,Halzen,Griffiths}. In other words, the spectrum of the DO (or particle/fermion) is given by $E_{DO}=E^+_{n,\kappa}$, while the spectrum of the anti-DO (or antiparticle/antifermion) is given by $E_{anti-DO}=-E_{n,\kappa}^-=+\vert E_{n,\kappa}\vert>0$, or better, $E_{DO}=E_{particle\ with\ positive\ energy}>0$, and $E_{anti-DO}=-E_{particle\ with\ negative\ energy}>0$ (i.e., a particle with negative energy is actually/(re)interpreted as an antiparticle with positive energy) \cite{Oli11,Strange,Greiner,Thomson,Halzen,Griffiths}. Therefore, both the DO and the anti-DO have positive energies as well as equal values ($E_{DO}=E_{anti-DO}>0$ or $\vert E^+_{n,\kappa}\vert=\vert E^-_{n,\kappa}\vert>0$), which implies that the spectrum is symmetric (around $E=0$), i.e., the spectrum obeys the condition: $E^+=-E^-$ or $E^-=-E^+$ (we will soon see such symmetry in the graphs). In particular, this ``energetic symmetry'' (or ``energy equality'') emphasizes the equilibrium between the DO and the anti-DO in the system (however, DO $\neq$ anti-DO, since they are Dirac fermions, i.e., particles that are not their own antiparticles). In fact, if we had any term outside the square root, then we would have an asymmetrical spectrum, i.e., $E_{DO}\neq E_{anti-DO}$ or $E^+\neq -E^-$. Besides, according to the Feynman-Stückelberg interpretation (a conceptual framework within quantum field theory (QFT) that physically reinterprets the energy-negative solutions/states), the negative-energy particles moving/going backward in spacetime (or for the past) are seen/(re)interpreted as positive-energy antiparticles moving/going forward in spacetime (or for the future), i.e., negative-energy particle solutions of the DE are (re)interpreted as being positive-energy antiparticle solutions \cite{Strange,Greiner,Thomson,Halzen,Griffiths}. Furthermore, in the absence of the DO ($\omega=0$), where the spectrum results in the rest energy $E=E_{rest}^{\pm}=\pm m_0/\alpha_s$ (``curved or effective rest energy'') for a free particle/antiparticle. Still regarding this rest energy (first term within the square root), we see that it is only affected by the original metric, i.e., the modified metric does not change the rest energy at all. However, the effect/influence of $\alpha_s$ and $\beta_s$ on the second term within the square root is equal, i.e., it does not matter if it is the original or modified metric; the second term will be the same. Consequently, this implies that the energies for the case of the original metric are greater than for the case of the modified metric ($E_{n,\kappa}^{original}>E_{n,\kappa}^{modified}$). So, as mentioned earlier, here, we must have $a<1$, which implies that $\alpha_s$ and $\beta_s$ are real parameters (i.e., we really have physically/quantumly consistent bound states).

So, we see that the spectrum \eqref{spectrum} is quantized in terms of the radial quantum number $n$ and the angular quantum number $\kappa$, and explicitly depends on the angular frequency $\omega$, curvature parameter $a$ (or parameters $\alpha_s$ and $\beta_s$), and on curved or effective rest mass $m_{\text{eff}}=m_0/\alpha_s$ or even, a curvature-dependent rest mass, or simply, a curvature-dependent mass), respectively. That is, in SI units, the effective rest energy is given by $E^\pm_{rest}=\pm m_{\text{eff}}c^2$. Besides, for $\kappa=l>0$ ($j=l-1/2$) as well as for $\kappa=-l-1<0$ ($j=l+1/2$), we have
\begin{equation}\label{spectrum4}
E^{original}_{n,\kappa}=\begin{cases}
\pm\sqrt{\frac{m^2_0}{\alpha^2_+}+4m_0\omega\left[n+\frac{\sqrt{\alpha^2_++4(l+1)(l+1-\alpha_+)}-2(l+1)}{4\alpha_+}+\frac{1}{4}\right]}, \ \ \ (j=l+1/2), 
\\
\pm\sqrt{\frac{m^2_0}{\alpha^2_+}+4m_0\omega\left[n+\frac{\sqrt{\alpha^2_++4l(l+\alpha_+)}+2l}{4\alpha_+}+\frac{1}{4}\right]}, \ \ \ \ \ \ \ \ \ \ \ \ \ \ \ \ (j=l-1/2),
\end{cases}
\end{equation}
\begin{equation}\label{spectrum5}
E^{modified}_{n,\kappa}=\begin{cases}
\pm\sqrt{m^2_0+4m_0\omega\left[n+\frac{\sqrt{\beta^2_-+4(l+1)(l+1-\beta_-)}-2(l+1)}{4\beta_-}+\frac{1}{4}\right]}, \ \ (j=l+1/2), 
\\
\pm\sqrt{m^2_0+4m_0\omega\left[n+\frac{\sqrt{\beta^2_-+4l(l+\beta_-)}+2l}{4\beta_-}+\frac{1}{4}\right]}, \ \ \ \ \ \ \ \ \ \ \ \ \ \ \ (j=l-1/2),
\end{cases}
\end{equation}
or better
\begin{equation}\label{spectrum6}
E^{original}_{n,\kappa}=\begin{cases}
\pm\sqrt{\frac{m^2_0}{\alpha^2_+}+4nm_0\omega}, \ \ \ \ \ \ \ \ \ \ \ \ \ \ \ \ \ \ \ \ \ \ (j=l+1/2), 
\\
\pm\sqrt{\frac{m^2_0}{\alpha^2_+}+4m_0\omega\left[n+\frac{l}{\alpha_+}+\frac{1}{2}\right]}, \ \ \ \ \ (j=l-1/2),
\end{cases}
\end{equation}
\begin{equation}\label{spectrum7}
E^{modified}_{n,\kappa}=\begin{cases}
\pm\sqrt{m^2_0+4nm_0\omega}, \ \ \ \ \ \ \ \ \ \ \ \ \ \ \ \ \ \ \ \ \ (j=l+1/2), 
\\
\pm\sqrt{m^2_0+4m_0\omega\left[n+\frac{l}{\beta_-}+\frac{1}{2}\right]}, \ \ \ \ \ (j=l-1/2),
\end{cases}
\end{equation}
or yet
\begin{equation}\label{spectrum8}
E^{original}_{n,\kappa}=\begin{cases}
\pm\sqrt{\frac{m^2_0}{\alpha^2_+}+m_0\omega[2N-2j+1]}, \ \ \ \ \ \ \ \ \ \ \ \ \ \ \ \ \ \ \ \ (j=l+1/2), 
\\
\pm\sqrt{\frac{m^2_0}{\alpha^2_+}+m_0\omega[2N_{\alpha_+}+2j_{\alpha_+}+3]}, \ \ \ \ \ (j_{\alpha_+}=l/\alpha_+-1/2),
\end{cases}
\end{equation}
\begin{equation}\label{spectrum9}
E^{modified}_{n,\kappa}=\begin{cases}
\pm\sqrt{m^2_0+m_0\omega[2N-2j+1]}, \ \ \ \ \ \ \ \ \ \ \ \ \ \ \ \ \ \ (j=l+1/2), 
\\
\pm\sqrt{m^2_0+m_0\omega[2N_{\beta_-}+2j_{\beta_-}+3]}, \ \ \ \  (j_{\beta_-}=l/\beta_--1/2),
\end{cases}
\end{equation}
where we define $N=2n+l$, $N_{\alpha_+}=2n+l/\alpha_+$, and $N_{\beta_-}=2n+l/\beta_-$ (i.e., three ``effective quantum numbers''. In fact, a more appropriate name for $N$ would be total quantum number since it depends on all the others (and is not influenced by curvature), and $N_{\alpha_s}$ would be curvature-dependent quantum numbers). Besides, the spectrum for $j=l-1/2$ (or $\kappa>0$) can be ``interpreted/seen'' as being the spectrum of the DO with spin down ($\downarrow$), that is, $E^\downarrow_{n,\kappa>0}=E^\downarrow_{n,l}$, and the spectrum for $j=l+1/2$ (or $\kappa<0$) as being the spectrum of the DO with spin up ($\uparrow$), that is, $E^\uparrow_{n,\kappa<0}=E^\uparrow_{n}$. In other words, for $\kappa>0$ (or spin down), the spectrum is quantized in terms of the quantum numbers $n$ and $l$, while for $\kappa<0$ (or spin up), it is quantized only in terms of $n$, respectively. Therefore, this implies that the energies for $\kappa>0$ are greater than for $\kappa<0$ (i.e., $\vert E^\pm_{n,\kappa>0}\vert>\vert E^\pm_{n,\kappa<0}\vert$, or yet $\vert E_{n,l}^{\downarrow}\vert>\vert E_{n}^{\uparrow}\vert$). It is interesting to highlight that in the case of the modified metric for $\kappa<0$, the spectrum does not depend on the parameter $\beta_-$ (see \eqref{spectrum7}), that is, it is as if the DO ``lived'' in the Minkowski spacetime. So, analogous to what happens with the DO or DE \cite{Bakke2,Oli9,Oli10,Oli11,Oli12,Oli13,Oli14} (and even with the Klein-Gordon equation \cite{Carvalho}) in the cosmic string spacetime (or Bonnor-Melvin-Lambda spacetime), where the angular momentum (or total magnetic quantum number) is modified by the topological/curvature parameter, here, this also happens, given by $j_{\alpha_+}$ (or $l/\alpha_+$) and $j_{\beta_-}$ (or $l/\beta_-$). Consequently, the spectra \eqref{spectrum6} and \eqref{spectrum7} for $j=l-1/2 $ ($\kappa>0$) also has its degeneracy broken (``undefined'' or not ``well-defined''), that is, the degeneracy of energy levels is broken due to the presence of the parameters $\alpha_+$ and $\beta_-$ (``tied/linked'' with $l$).

In particular, in the absence of the cloud of strings ($\alpha_s=\beta_s=+1$), we recover exactly the usual spectrum of the DO in the $(3+1)$-dimensional Minkowski spacetime (as it should be), given as follows \cite{Moshinsky,Strange}
\begin{equation}\label{usualspectrum}
E^\pm_{n,j}=\begin{cases}
\pm\sqrt{m^2_0+m_0\omega\left[2N-2j+1\right]}, \ \ \ \ (j=l+1/2),
\\
\pm\sqrt{m^2_0+m_0\omega\left[2N+2j+3\right]}, \ \ \ \ (j=l-1/2),
\end{cases}
\end{equation}
where the spectrum has an infinite degeneracy for $j=l+1/2$ (i.e., the spectrum depends only on $n$), and a finite degeneracy for $j=l-1/2$ (i.e., the spectrum depends on both $n$ and $l$), respectively. However, it is important to highlight that the $N's$ in \eqref{usualspectrum} are different (since $j's$ are different), that is, the first $N$ has values $0,1,2,\ldots$ (since $l=0,1,2,\ldots$), while the second has values $1,2,3,\ldots$ (since $l=1,2,3,\ldots$).

Now, let us graphically analyze the behavior of the relativistic spectrum as a function of the angular frequency $\omega$ and of the curvature parameter $a$ for three different values of the quantum numbers $n$ and $\kappa$. So, since we want to investigate the influence of $a$ on both the original and modified spectra, here, we will work only with the spectra for $\kappa>0$ ($\kappa=j+1/2=l=1,2,3,\ldots$), given by \eqref{spectrum6} and \eqref{spectrum7}. That is, we chose this because such spectra provide the maximum energies for the system (where we use $l$ instead of $\kappa$ for convenience). In this way, the values of the quantum numbers $n$ and $l$ allow us to analyze the behavior of the spectrum in two different cases/scenarios, which are: while $n$ varies ($n=0,1,2$), $l$ remains fixed ($l=1$), and while $l$ varies ($l=1,2,3$), $n$ remains fixed ($n=0$), respectively. Furthermore, for simplicity and convenience, we will assume that we have a ``unit mass'', i.e., $m_0=1$ (in fact, we use this since here we are already using the natural unit system where $\hbar=c=1$). Therefore, the spectra that will be used to plot the graphs are given by (in order to see the symmetry of the energy levels about $E=0$, we carry both signs $\pm$. However, it is always good to remember that the physical energies of the anti-DO are also positive)
\begin{equation}\label{spectra}
E^\pm_{n,l}(\omega,a)=\begin{cases}
E^\pm_{original}(\omega,a)=\pm\sqrt{\frac{1}{(1-a)}+4\omega\left[n+\frac{l}{\sqrt{1-a}}+\frac{1}{2}\right]},
\\
E^\pm_{modified}(\omega,a)=\pm\sqrt{1+4\omega\left[n+\frac{l}{\sqrt{1-a}}+\frac{1}{2}\right]}.
\end{cases}
\end{equation}

\begin{figure}[ht]
    \centering
    \subfigure{%
        \includegraphics[width=0.49\textwidth]{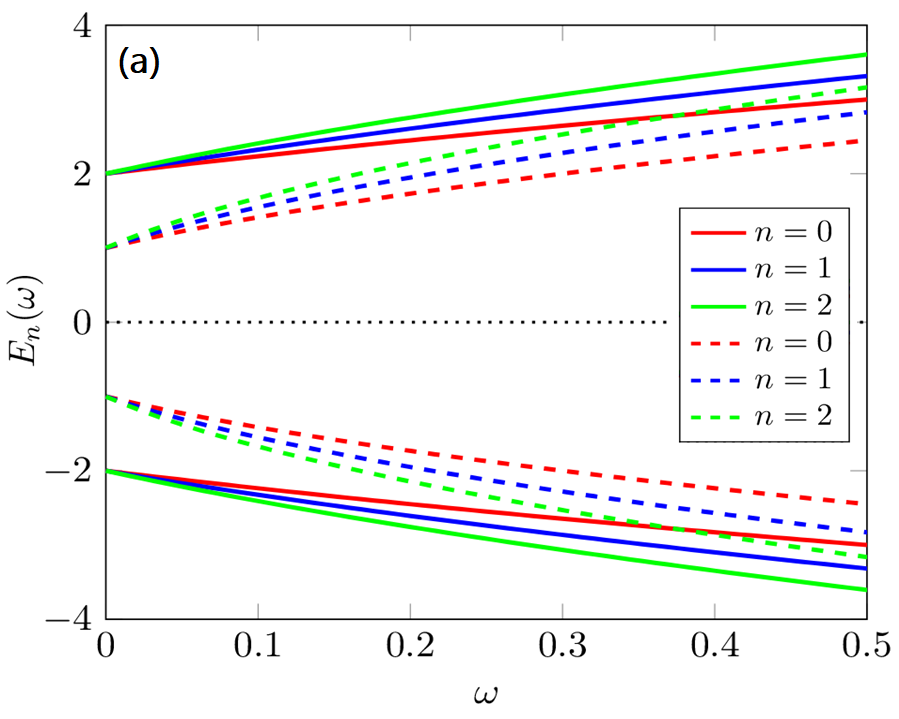}
    }\hfill
    \subfigure{%
        \includegraphics[width=0.49\textwidth]{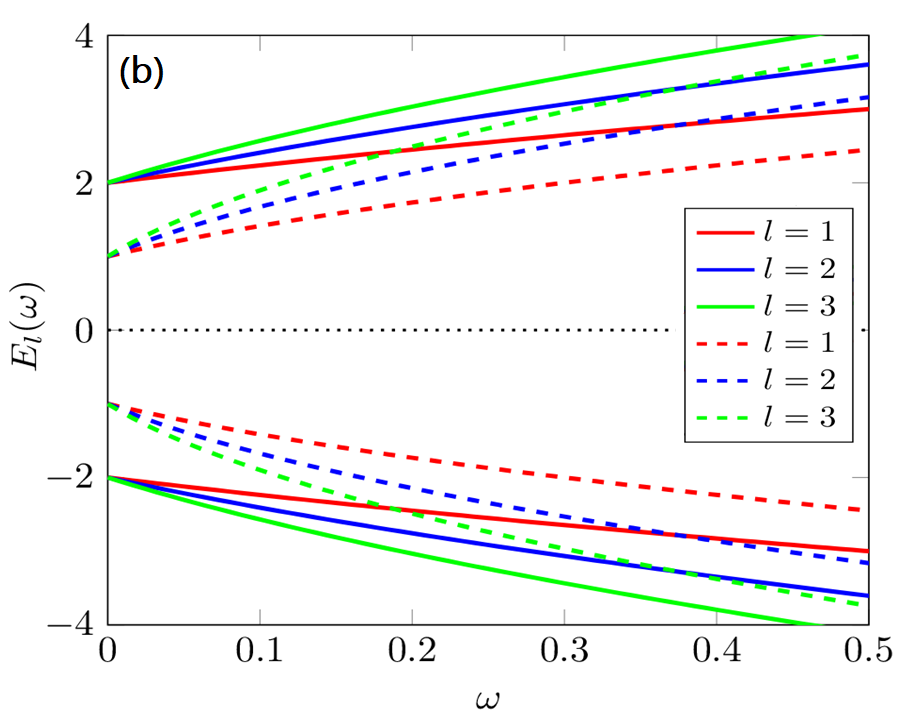}
    }
    \caption{Behavior of $E_n(\omega)$ vs. $\omega$ for three different values of $n$ with $l=1$ (a) and the behavior of $E_l(\omega)$ vs. $\omega$ for three different values of $l$ with $n=0$ (b), where the solid curves are for the original metric and the dashed curves for the modified metric.}
    \label{fig1}
\end{figure}

So, in Fig. \ref{fig1}, we have the behavior of $E_{n}(\omega)$ vs. $\omega$ for $n=0,1,2$ with $l=1$ (a) and the behavior of $E_{l}(\omega)$ vs. $\omega$ for $l=1,2,3$ with $n=0$ (b), where the solid curves are for the original metric and the dashed curves for the modified metric (we use $a=0.75$). According to this figure, we can extract some valuable/interesting information. For example, we see that the behavior of the two graphs is very similar, i.e., has the same profile/appearance (basically, the difference is only in the energy values, but except for $n=0$ and $l=1$, which appear in both graphs and are described by the red solid and dashed curves). In fact, this happens because the quantum numbers $n$ and $l$ (assuming a fixed value for $1/\sqrt{1-a}$) act in a similar way within the square root (where both have positive values). Therefore, in both graphs (for the DO and anti-DO), the energies increase with increasing $n$ and $l$ (as it must be to be physically/quantumly consistent; otherwise, something would be wrong). Consequently, this implies that the energy difference (or the spacing) between two consecutive levels is positive (i.e., $\Delta E_{n,l}=E_{n+1,l+1}-E_{n,l}>0$). In particular, this energy difference increases as a function of $\omega$ (i.e., $\Delta E_{n,l}(\omega+\delta\omega)>\Delta E_{n,l}(\omega)$ or $\Delta E_{n,l}(\omega=0.5)>\Delta E_{n,l}(\omega=0.4)>\Delta E_{n,l}(\omega=0.3)\ldots$), that is, the larger $\omega$, the greater the spacing between the energy levels (in fact, this happens because here the energy curves are not parallel to each other, but rather they move away from each other as $\omega$ increases). Besides, the energies (for a given value of $n$ or $l$) also increase as a function of $\omega$, that is, $\omega$ has the objective of increasing the energies (as well as the spacing between them, as we mentioned above). Consequently, this implies that the variation of energy as a function of $\omega$ is positive (i.e., $\Delta E (\omega)=E(\omega+\delta\omega)-E(\omega)=E_{final}(\omega)-E_{initial}(\omega)>0$). Here, it is important to highlight that $\Delta E_{n,l}$ (or even $\Delta E_{n,l}(\omega)$) is not the same thing as $\Delta E(\omega)$, that is, they are different things (have conceptually different meanings). In fact, in quantum mechanics, $\Delta E_{n,l}$ refers to the (net) result of the difference/transition between the energies of two different quantum states (or simply the spacing between two quantum states or energy levels), while $\Delta E(\omega)$ refers to the change in energy (of the same quantum state or energy level) for a given change in $\omega$ (or simply the variation of the energy for a given variation of $\omega$, which can be small or large). In short, the energy difference is a property/characteristic between two different quantum states, while the energy variation is a physical process or a change in the system's energy (with respect to some parameter/variable of the system). Besides, as we said earlier (and can be easily seen in the graphs), the energies for the original metric are greater than for the modified metric. For the sake of curiosity, for $\omega=0$, we have $E^\pm_{original}(\omega)=\pm 2$ and $E^\pm_{modified}(\omega)=\pm 1$ (this can be easily seen in Fig. \ref{fig1} or in \eqref{spectra}), that is, the energy for the original metric is twice the energy for the modified metric ($E^\pm_{original}(\omega)=2E^\pm_{modified}(\omega)$).

\begin{figure}[ht]
    \centering
    \subfigure{%
        \includegraphics[width=0.49\textwidth]{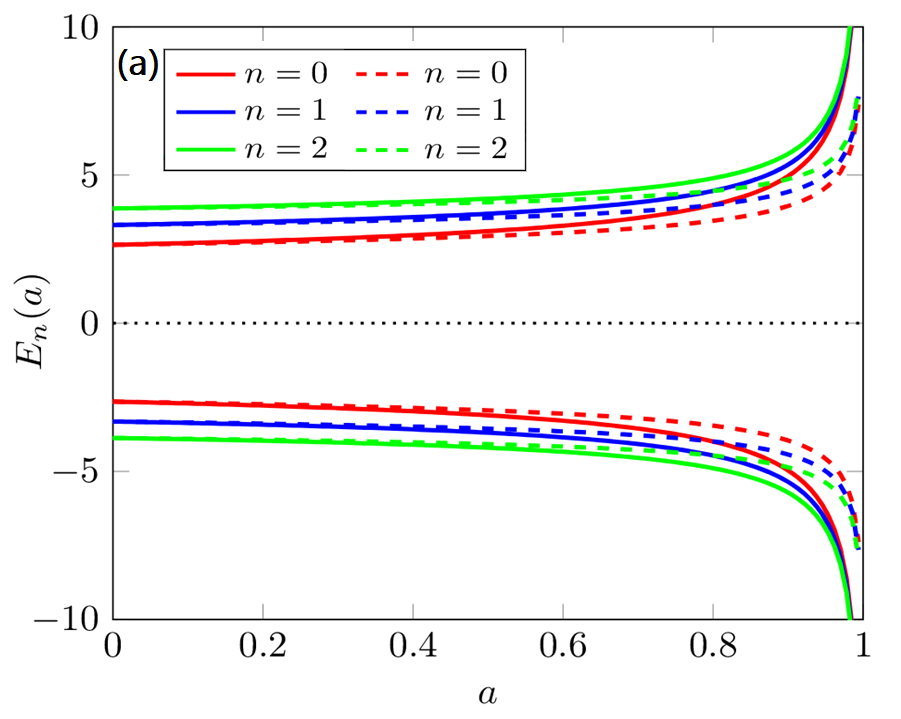}
    }\hfill
    \subfigure{%
        \includegraphics[width=0.49\textwidth]{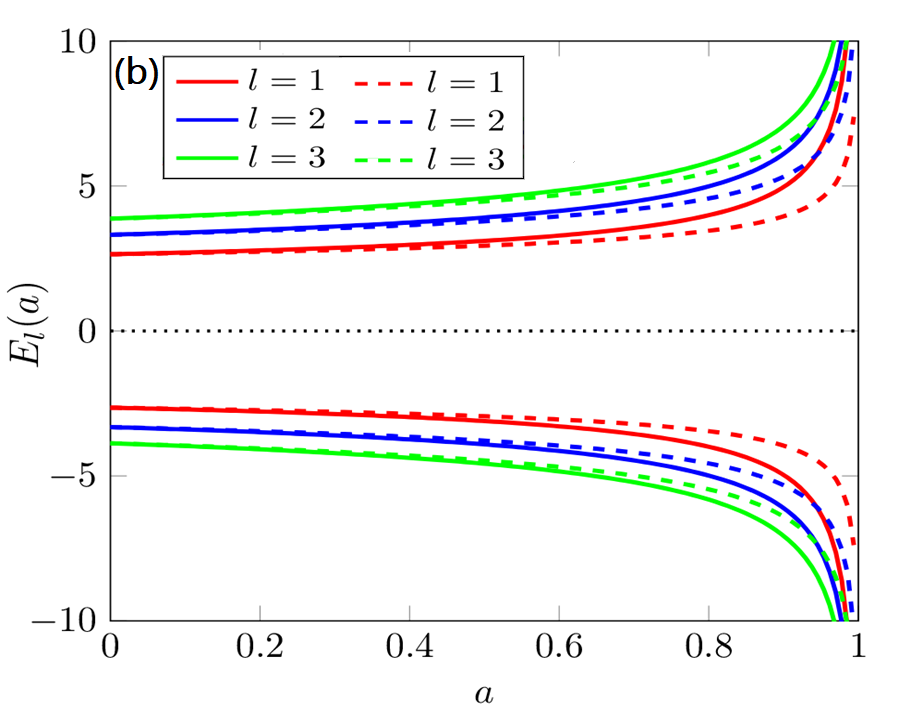}
    }
    \caption{Behavior of $E_n(a)$ vs. $a$ for three different values of $n$ with $l=1$ (a) and the behavior of $E_l(a)$ vs. $a$ for three different values of $l$ with $n=0$ (b), where the solid curves are for the original metric and the dashed curves for the modified metric.}
    \label{fig2}
\end{figure}

Already in Fig. \ref{fig2}, we have the behavior of $E_{n}(a)$ vs. $a$ for $n=0,1,2$ with $l=1$ (a) and the behavior of $E_{l}(a)$ vs. $a$ for $l=1,2,3$ with $n=0$ (b), where the solid curves are for the original metric and the dashed curves for the modified metric (we use $\omega=1$). According to this figure, we can also extract some valuable/interesting information. So, analogous to the previous figure, here, we see that the behavior of the two graphs is also very similar, i.e., has the same profile/appearance (basically, the difference is only in the energy values, but except for $n=0$ and $l=1$, which appear in both graphs and are described by the red solid and dashed curves). Therefore, in both graphs (for the DO and anti-DO), the energies increase with increasing $n$ and $l$ (as it must be to be physically/quantumly consistent); however, they tend to ``come together'' (or have the same value) for $a\to 1$ (i.e., the greater the curvature of spacetime, the closer the energies will be to each other). In that way, the energy difference (or the spacing) between two consecutive levels is positive for $0\leq a<1$ ($\Delta E_{n,l}>0$) and tends to zero for $a\to 1$ ($\Delta E_{n,l}=0$), respectively. In particular, this energy difference is practically/approximately constant for $0\leq a \lesssim 0.5$ (i.e., almost evenly spaced or practically parallel curves), and decreases for $a>0.5$ (i.e., increasingly smaller spacing), in which it tends to zero for $a\to 1$ (i.e., the spacing tending to zero). In other words, this is a consequence of the fact that the energies increase very little for $0\leq a \lesssim 0.5$ (the energies for the two metrics are practically the same) and a lot as $a$ increases (tending to infinity for $a\to 1$). That is, we have $\Delta E_{n,l}(0\leq a \lesssim 0.5)\cong constant$ (or $\Delta E_{n,l}(a=0.0)\cong\Delta E_{n,l}(a=0.1)\cong\Delta E_{n,l}(a=0.2)\ldots\cong\Delta E_{n,l}(a=0.5)\cong constant$), and $\Delta E_{n,l}(a\to 1)=0$. Therefore, the energies increase as a function of $a$; however, this increase is only much more significant for increasingly larger values of $a$ (i.e., with the increase in the curvature of spacetime). Consequently, this implies that the variation of energy as a function of $a$ is practically/approximately zero for $0\leq a \lesssim 0.5$ ($\Delta E(a)=E(a+\delta a)-E(a)\cong 0$), and positive for $a>0.5$ ($\Delta E(a)>0$), in which it tends to infinity for $a\to 1$ ($\Delta E(a\to 1)\to\infty$). Besides, as we said earlier (and can be easily seen in the graphs), the energies for the original metric are greater than for the modified metric (however, this only starts to become more significant for $a>0.5$).

Now, let us focus our attention on the form of the original Dirac spinor and later on its normalization; that is, here, we will obtain the normalized Dirac spinor. Indeed, for a system to be completely described within relativistic quantum mechanics (and nonrelativistic quantum mechanics), the Dirac spinor must be normalizable (or normalized) to ensure that the probabilities associated with the measurement results are finite and well-defined. Then, since we already have the form of the upper radial component, given by \eqref{u2}, we now need to find the form of the lower radial component, given by $v(r)$. To do this, we can simply start directly from the upper radial component $u(r)$, or better, from $u(y)$ \cite{Bragança,Oli11}. In fact, using $y=m_0\omega r^2$, we can easily return to the original variable $r$ later. Thus, writing Eq. \eqref{EDO8} in terms of the variable $y$, we have
\begin{equation}\label{EDO12}
\left(E+\frac{m_0}{\alpha_s}\right)v(y)-\sqrt{m_0\omega}\left(2\sqrt{y}\frac{d}{dy}+\sqrt{y}+\frac{\kappa}{\alpha_s\beta_s\sqrt{y}}\right)u(y)=0,
\end{equation}
or better
\begin{equation}\label{v1}
v(y)=\frac{\sqrt{m_0\omega}}{\left(E+\frac{m_0}{\alpha_s}\right)}\left(2\sqrt{y}\frac{d}{dy}+\sqrt{y}+\frac{\kappa}{\alpha_s\beta_s\sqrt{y}}\right)u(y).
\end{equation}

Therefore, substituting \eqref{u2} (with $\mu-\nu+1/2=-n$) into \eqref{v1}, we obtain
\begin{equation}\label{v2}
v(y)=\frac{C(m_0\omega)^{3/4}y^{\mu-1/4}e^{-y/2}}{2\left(E+\frac{m_0}{\alpha_s}\right)}\left[\left(1+\frac{2\kappa}{\alpha_s\beta_s}+4\mu\right){_{1}F_{1}}(-n,2\mu+1;y)-\frac{4ny}{(2\mu+1)}{_{1}F_{1}}(-n+1,2\mu+2;y)\right],
\end{equation}
or in terms of the original variable $r$, such as
\begin{equation}\label{v3}
v(r)=\xi(r)\left[\left(1+\frac{2\kappa}{\alpha_s\beta_s}+4\mu\right){_{1}F_{1}}(-n,2\mu+1;m_0\omega r^2)-\frac{4nm_0\omega r^2}{(2\mu+1)}{_{1}F_{1}}(-n+1,2\mu+2;m_0\omega r^2)\right],
\end{equation}
where we define for convenience
\begin{equation}\label{xi}
\xi(r)=\frac{C(m_0\omega)^{3/4}(\sqrt{m_0\omega}r)^{2\mu-1/2}e^{-m_0\omega r^2/2}}{2\left(E+\frac{m_0}{\alpha_s}\right)}.
\end{equation}

However, in order to obtain a normalized spinor later, that is, to find the normalization constant of the system, it is more advisable to work directly with the associated Laguerre polynomials rather than the confluent hypergeometric functions \cite{Bragança,Oli11}. In particular, this is due to the fact that the normalization condition depends on the integral of the probability density (and this, in turn, depends on the spinor and its adjoint or Hermitian conjugate). Consequently, working directly with the associated Laguerre polynomials is much more appropriate/advantageous than with the confluent hypergeometric functions \cite{Bragança,Oli11}. So, knowing that the confluent hypergeometric functions are written in terms of the associated Laguerre polynomials as follows \cite{Bragança,Arfken}
\begin{equation}
_{1}F_{1}(-n,b+1;x)=\frac{n!b!}{(n+b)!}L^{b}_n (x),
\end{equation}
implies that \eqref{u2} and \eqref{v2} will be written as
\begin{eqnarray}
&& u(y)=\frac{n!(2\mu)!C(m_0\omega)^{1/4}}{(n+2\mu)!}y^{\mu+1/4}e^{-y/2}L^{2\mu}_{n}(y),
\label{u4}\\
&& v(y)=\frac{C(m_0\omega)^{3/4}y^{\mu-1/4}e^{-y/2}}{2\left(E+\frac{m_0}{\alpha_s}\right)}\left[\frac{n!(2\mu)!}{(n+2\mu)!}\left(1+\frac{2\kappa}{\alpha_s\beta_s}+4\mu\right)L^{2\mu}_{n}(y)-\frac{4ny}{(2\mu+1)}\frac{(n-1)!(2\mu+1)!}{(n+2\mu)!}L^{2\mu+1}_{n-1}(y)\right],
\label{v4}
\end{eqnarray}
or in terms of $r$, \eqref{u3} and \eqref{v3} will be written as
\begin{eqnarray}
&& u(r)=\frac{n!(2\mu)!C(m_0\omega)^{1/4}}{(n+2\mu)!}(\sqrt{m_0\omega}r)^{2\mu+1/2}e^{-m_0\omega r^2/2}L^{2\mu}_{n}(m_0\omega r^2),
\label{u5}\\
&& v(r)=\xi(r)\left[\frac{n!(2\mu)!}{(n+2\mu)!}\left(1+\frac{2\kappa}{\alpha_s\beta_s}+4\mu\right)L^{2\mu}_{n}(m_0\omega r^2)-\frac{4n!(2\mu)!\delta_{n,n\geq 1}m_0\omega r^2}{(n+2\mu)!}L^{2\mu+1}_{n-1}(m_0\omega r^2)\right].
\label{v5}
\end{eqnarray}
where we use the fact that $\frac{(2\mu+1)!}{(2\mu+1)}=(2\mu)!$ and $n(n-1)!=n!$. In addition, it is also important to mention that since the derivative of $L^b_n (x)$ for $n=0$ (ground state) is zero ($\frac{dL^b_0 (x)}{dx}=0$), it implies that the second term of \eqref{v4} or \eqref{v5} for $n=0$ is also zero (i.e., undefined). Consequently, such terms are valid (or nonzero) only for $n\geq 1$. So, we did $n!\to n!\delta_{n,n\geq 1}$ in the second term (this Kronecker delta is also used in \cite{Bragança}).

Therefore, using \eqref{u5} and \eqref{v5}, the spinor \eqref{spinor} takes the following form
\begin{equation}\label{spinor2}
\Phi=e^{-iEt}\left(
           \begin{array}{c}
           \frac{n!(2\mu)!C(m_0\omega)^{1/4}}{(n+2\mu)!}(\sqrt{m_0\omega}r)^{2\mu+1/2}e^{-m_0\omega r^2/2}L^{2\mu}_{n}(m_0\omega r^2)\chi^{m_j}_{\kappa}(\theta,\phi)\\
           i\xi(r)\left[\frac{n!(2\mu)!}{(n+2\mu)!}\left(1+\frac{2\kappa}{\alpha_s\beta_s}+4\mu\right)L^{2\mu}_{n}(m_0\omega r^2)-\frac{4n!(2\mu)!\delta_{n,n\geq 1}m_0\omega r^2}{(n+2\mu)!}L^{2\mu+1}_{n-1}(m_0\omega r^2)\right]\chi^{m_j}_{-\kappa}(\theta,\phi)\\
           \end{array}
         \right),
\end{equation}
or yet
\begin{equation}\label{spinor3}
\Phi=e^{-iEt}\left(
           \begin{array}{c}
           \frac{n!(2\mu)!C(m_0\omega)^{\mu+1/2}}{(n+2\mu)!}r^{2\mu+1/2}e^{-m_0\omega r^2/2}L^{2\mu}_{n}(m_0\omega r^2)\chi^{m_j}_{\kappa}(\theta,\phi)\\
           i\xi(r)\left[\frac{n!(2\mu)!}{(n+2\mu)!}\left(1+\frac{2\kappa}{\alpha_s\beta_s}+4\mu\right)L^{2\mu}_{n}(m_0\omega r^2)-\frac{4n!(2\mu)!\delta_{n,n\geq 1}m_0\omega r^2}{(n+2\mu)!}L^{2\mu+1}_{n-1}(m_0\omega r^2)\right]\chi^{m_j}_{-\kappa}(\theta,\phi)\\
           \end{array}
         \right),
\end{equation}
where
\begin{equation}\label{xi2}
\xi(r)=\frac{C(m_0\omega)^{\mu+1/2}r^{2\mu-1/2}e^{-m_0\omega r^2/2}}{2\left(E+\frac{m_0}{\alpha_s}\right)}.
\end{equation}

In that way, using \eqref{psi} and knowing that $\Psi_D=\Psi_D (r,\theta,\phi,t)=S\psi$, where $S$ is given in \eqref{S2}, implies that the (original) Dirac spinor takes the following form
\begin{equation}\label{spinor4}
\Psi_D=\frac{Se^{-iEt}}{r\sqrt{\sin\theta}}\left(
           \begin{array}{c}
           \frac{n!(2\mu)!C(m_0\omega)^{\mu+1/2}}{(n+2\mu)!}r^{2\mu+1/2}e^{-m_0\omega r^2/2}L^{2\mu}_{n}(m_0\omega r^2)\chi^{m_j}_{\kappa}(\theta,\phi)\\
           i\xi(r)\left[\frac{n!(2\mu)!}{(n+2\mu)!}\left(1+\frac{2\kappa}{\alpha_s\beta_s}+4\mu\right)L^{2\mu}_{n}(m_0\omega r^2)-\frac{4n!(2\mu)!\delta_{n,n\geq 1}m_0\omega r^2}{(n+2\mu)!}L^{2\mu+1}_{n-1}(m_0\omega r^2)\right]\chi^{m_j}_{-\kappa}(\theta,\phi)\\
           \end{array}
         \right).
\end{equation}

So, to find the normalization constant $C$, we have to use the normalization condition for the Dirac spinor in curved spacetimes (or ``curved spinor inner product''). For example, according to Refs. \cite{Bragança,Oli11,Antoine,Belgiorno,Iyer,Gerbert,Leclerc}, the normalization condition for the Dirac spinor in curved spacetimes is given by the following volume integral in spherical coordinates (for bound states, the total probability must be 1, i.e., the total probability of finding the DO somewhere in space is 100\%)
\begin{equation}\label{integral1}
\langle \Psi_D | \Psi_D\rangle=\int_{t=const.} J^t (x)dV_{curved}=\int\bar{\Psi}_D\gamma^t (x)\Psi_D \sqrt{-g(x)}d^3x=\int_{0}^{\infty}\int_{0}^{2\pi}\int_{0}^{\pi}\sqrt{-g(x)}\bar{\Psi}_D\gamma^t (x)\Psi_D drd\theta d\phi=1,
\end{equation}
where $J^t (x)=\rho(x)\geq 0$ is the curved probability density (a positive definite quantity), $dV_{curved}=\sqrt{-g(x)}d^3 x=\sqrt{-g(x)}dr d\theta d\phi$ is the curved volume element (for $t$=constant), $g(x)$ is the determinant of the metric ($g(x)$=det$g_{\mu\nu}(x)<0$ or $g(x)=-\vert g(x) \vert$), $\bar{\Psi}_D=\Psi_D^\dagger \gamma^0$ is the adjoint spinor (or Dirac adjoint), $\Psi^\dagger$ denotes the Hermitian conjugate (complex conjugate transpose) of $\Psi_D=(\Psi_1,\Psi_2)^T$ (i.e., $\Psi^\dagger=(\Psi_1^\star,\Psi_2^\star)$), and $\gamma^t (x)=e^t_{\ a}(x)\gamma^a$ is the zero/temporal component of the curved gamma matrices $\gamma^\mu (x)$, respectively. On the other hand, we can still write $\sqrt{-g(x)}$ in terms of (inverse) tetrads, that is: $\sqrt{-g(x)}$=det$(e^a_{\ \mu}(x))$ \cite{Lawrie}. In this way, we have
\begin{equation}\label{integral2}
\int_{0}^{\infty}\int_{0}^{2\pi}\int_{0}^{\pi}det(e^a_{\ \mu}(x))\bar{\Psi}_D\gamma^t(x)\Psi_D drd\theta d\phi=1,
\end{equation}
or (with $\gamma^t(x)=e^t_{\ 0}(x)\gamma^0$)
\begin{equation}\label{integral3}
\int_{0}^{\infty}\int_{0}^{2\pi}\int_{0}^{\pi}det(e^a_{\ \mu}(x))e^t_{\ 0}(x)\Psi^{\dagger}_D \Psi_D drd\theta d\phi=1,
\end{equation}
or better (using \eqref{tetrads})
\begin{equation}\label{integral4}
\frac{\beta_s^2}{\alpha_s}\int_{0}^{\infty}\int_{0}^{2\pi}\int_{0}^{\pi}\Psi^{\dagger}_D \Psi_D r^2\sin{\theta}drd\theta d\phi=1.
\end{equation}

Knowing that $\Psi_D=S\psi=\frac{S\Phi}{r\sqrt{\sin{\theta}}}$, we have
\begin{equation}\label{integral5}
\frac{\beta_s^2}{\alpha_s}\int_{0}^{\infty}\int_{0}^{2\pi}\int_{0}^{\pi}\Phi^{\dagger} \Phi drd\theta d\phi=1,
\end{equation}
or yet (using \eqref{spinor})
\begin{equation}\label{integral6}
\frac{\beta_s^2}{\alpha_s}\int_{0}^{\infty}\int_{0}^{2\pi}\int_{0}^{\pi}\left\{[u(r)\chi^{m_j}_{\kappa}(\theta,\phi)]^2+[v(r)\chi^{m_j}_{-\kappa}(\theta,\phi)]^2\right\}drd\theta d\phi=1.
\end{equation}

Now, knowing/considering that the angular part of the spinor is already normalized \cite{Strange,Greiner}, implies that we will only have the normalization of the radial part of the spinor (``radial normalization condition''), given as follows
\begin{equation}\label{integral7}
\frac{\beta_s^2}{\alpha_s}\int_{0}^{\infty}\left[u^2(r)+v^2(r)\right]dr=1.
\end{equation}

In particular, for $\beta_s=+1$, we will have a normalization condition similar to that of the DO in the global monopole spacetime \cite{Bragança}. Already for $\alpha_s=\beta_s=+1$ (absence of the cloud of strings), we will have the normalization condition of the DE in the Minkowski spacetime in spherical coordinates \cite{Strange,Greiner,Schluter}. Besides, doing $\alpha_s=\beta_s=+1$ in \eqref{integral5}, we have the normalization condition of Ref. \cite{Schluter} (indeed, the spinor \eqref{spinor} is basically the spinor (5.42) of this reference without $r$ in the denominator and with the exponential $e^{-iEt}$).

So, our objective is to solve the integral \eqref{integral7}. However, it is convenient to write this integral in terms of the variable $y$. For example, knowing that $y=m_0\omega r^2$, implies that $dy=2m_0 \omega r dr=2\sqrt{m_0\omega}y^{1/2}dr\to dr=\frac{y^{-1/2}}{2\sqrt{m_0\omega}}dy$. Therefore, we have
\begin{equation}\label{integral8}
\frac{\beta_s^2}{2\alpha_s\sqrt{m_0\omega}}\int_{0}^{\infty}\left[u^2(y)+v^2(y)\right]y^{-1/2}dy=1.
\end{equation}

In that way, substituting \eqref{u4} and \eqref{v4} (with $(2\mu+1)!/(2\mu+1)=(2\mu)!$ and $n(n-1)!=n! \delta_{n,n\geq 1}$) in \eqref{integral8}, we obtain
\begin{equation}\label{integral9}
\frac{\beta_s^2}{2\alpha_s}\left[\frac{n!(2\mu)!C}{(n+2\mu)!}\right]^2\int_{0}^{\infty}\left\{y^{2\mu}e^{-y}[L^{2\mu}_{n}(y)]^2+\frac{m_0\omega y^{2\mu-1}e^{-y}}{4\left(E+\frac{m_0}{\alpha_s}\right)^2}\left[\left(1+\frac{2\kappa}{\alpha_s\beta_s}+4\mu\right)L^{2\mu}_{n}(y)-4\delta_{n,n\geq 1}yL^{2\mu+1}_{n-1}(y)\right]^2\right\}dy=1,
\end{equation}
or better
\begin{equation}\label{integral10}
\frac{\beta_s^2}{2\alpha_s}\left[\frac{n!(2\mu)!C}{(n+2\mu)!}\right]^2\left\{I_1+\frac{m_0\omega}{4\left(E+\frac{m_0}{\alpha_s}\right)^2}\left[\left(1+\frac{2\kappa}{\alpha_s\beta_s}+4\mu\right)^2 I_2-8\delta_{n,n\geq 1}\left(1+\frac{2\kappa}{\alpha_s\beta_s}+4\mu\right)I_3+[4\delta_{n,n\geq 1}]^2I_4\right]\right\}dy=1,
\end{equation}
where $I_1$, $I_2$, $I_3$ and $I_4$ are integrals, defined as follows
\begin{eqnarray}\label{integrals}
&& I_1=\int_{0}^{\infty}e^{-y}y^{2\mu}\left[L^{2\mu}_n (y)\right]^2 dy,
\\
&& I_2=\int_{0}^{\infty}e^{-y}y^{2\mu-1}\left[L^{2\mu}_n (y)\right]^2 dy,
\\
&& I_3=\int_{0}^{\infty}e^{-y}y^{2\mu}L^{2\mu}_n (y)L^{2\mu+1}_{n-1}(y) dy,
\\
&& I_4=\int_{0}^{\infty}e^{-y}y^{2\mu+1}\left[L^{2\mu+1}_{n-1}(y)\right]^2 dy.
\end{eqnarray}

However, we can obtain the solutions of the above integrals in two ways: by solving each integral individually/separately or via ``mapping''. For simplicity and without loss of generality, we will opt for the second way. So, doing $2\mu \to -Ms+1/2$ (with $y\to w$) in \eqref{integrals}, we obtain exactly the integrals of Ref. \cite{Oli11}. Therefore, by doing $-Ms+1/2 \to 2\mu$ on the solutions of the integrals of Ref. \cite{Oli11}, we obtain
\begin{eqnarray}\label{integrals3}
&& \int_{0}^{\infty}e^{-y}y^{2\mu}\left[L^{2\mu}_n (y)\right]^2 dy=\frac{(n+2\mu)!}{n!},
\\
&& \int_{0}^{\infty}e^{-y}y^{2\mu-1}\left[L^{2\mu}_n (y)\right]^2 dy=\frac{(n+2\mu)!}{n!2\mu},
\\
&& \int_{0}^{\infty}e^{-y}y^{2\mu}L^{2\mu}_n (y)L^{2\mu+1}_{n-1}(y) dy=-\frac{(n+2\mu+1)!}{n!(2\mu-2)},
\\
&& \int_{0}^{\infty}e^{-y}y^{2\mu+1}\left[L^{2\mu+1}_{n-1}(y)\right]^2 dy=\frac{(n+2\mu+1)!}{n!(2\mu+2)}.
\end{eqnarray}

Consequently, we obtain the following relativistic normalization constant for our problem
\begin{equation}\label{normalizationconstantofthesystem}
C=C_{n,\kappa}=\sqrt{
\frac{
  2\alpha_s(n+2\mu)! 
}{
 n![(2\mu)!]^2 \beta_s^2
  \left\{
    1+\frac{m_0\omega}{4\left(E+\frac{m_0}{\alpha_s}\right)^2}
    \left[
      \frac{\left(1+\frac{2\kappa}{\alpha_s\beta_s}+4\mu\right)^2}{2\mu}
      + \frac{4\delta_{n,n\geq 1}(n+2\mu+1)\left(1+\frac{2\kappa}{\alpha_s\beta_s}+4\mu\right)}{(\mu-1)}
      +\frac{8\delta_{n,n\geq 1}(n+2\mu+1)}{(\mu+1)}
    \right]
  \right\}
}
},
\end{equation}
or better
\begin{equation}\label{normalizationconstantofthesystem2}
C=\sqrt{
\frac{
  2\alpha_s(n+2\mu)! 
}{
 n![(2\mu)!]^2 \beta_s^2
  \left\{
    1+\frac{2m_0\omega}{\left(E+\frac{m_0}{\alpha_s}\right)^2}
    \left[
      \frac{\left(\frac{1}{2}+\frac{\kappa}{\alpha_s\beta_s}+2\mu\right)^2}{4\mu}
      +\delta_{n,n\geq 1}(n+2\mu+1)\left(\frac{\left(\frac{1}{2}+\frac{\kappa}{\alpha_s\beta_s}+2\mu\right)}{(\mu-1)}
      +\frac{1}{(\mu+1)}\right)
    \right]
  \right\}
}
}.
\end{equation}

Therefore, using the normalization constant \eqref{normalizationconstantofthesystem2} in \eqref{spinor4}, we have the following normalized Dirac spinor for the relativistic bound states of the DO in the curved spacetime of a cloud of strings
\begin{equation}\label{spinor5}
\Psi_D=\frac{\bar{C}Se^{-iEt}}{r\sqrt{\sin\theta}}\left(
           \begin{array}{c}
           r^{2\mu+1/2}e^{-m_0\omega r^2/2}L^{2\mu}_{n}(m_0\omega r^2)\chi^{m_j}_{\kappa}(\theta,\phi)\\
           \frac{i r^{2\mu-1/2}e^{-m_0\omega r^2/2}}{\left(E+\frac{m_0}{\alpha_s}\right)}\left[\left(\frac{1}{2}+\frac{\kappa}{\alpha_s\beta_s}+2\mu\right)L^{2\mu}_{n}(m_0\omega r^2)-2\delta_{n,n\geq 1}m_0\omega r^2 L^{2\mu+1}_{n-1}(m_0\omega r^2)\right]\chi^{m_j}_{-\kappa}(\theta,\phi)\\
           \end{array}
         \right),
\end{equation}
or yet
\begin{equation}\label{spinor6}
\Psi_D=\frac{\bar{C}Se^{-iEt}r^{2\mu-1/2}e^{-m_0\omega r^2/2}}{r\sqrt{\sin\theta}}\left(
           \begin{array}{c}
           rL^{2\mu}_{n}(m_0\omega r^2)\chi^{m_j}_{\kappa}(\theta,\phi)\\
           \frac{i}{\left(E+\frac{m_0}{\alpha_s}\right)}\left[\left(\frac{1}{2}+\frac{\kappa}{\alpha_s\beta_s}+2\mu\right)L^{2\mu}_{n}(m_0\omega r^2)-2\delta_{n,n\geq 1}m_0\omega r^2 L^{2\mu+1}_{n-1}(m_0\omega r^2)\right]\chi^{m_j}_{-\kappa}(\theta,\phi)\\
           \end{array}
         \right),
\end{equation}
where $\bar{C}$ is a ``new'' normalization constant, defined as follows
\begin{equation}\label{newnormalizationconstant}
\bar{C}=\frac{n!(2\mu)!(m_0\omega)^{\mu+1/2}C}{(n+2\mu)!}=\sqrt{
\frac{
  2n!\alpha_s (m_0\omega)^{2\mu+1}
}{
 (n+2\mu)!\beta_s^2
  \left\{
    1+\frac{2m_0\omega}{\left(E+\frac{m_0}{\alpha_s}\right)^2}
    \left[
      \frac{\left(\frac{1}{2}+\frac{\kappa}{\alpha_s\beta_s}+2\mu\right)^2}{4\mu}
      +\delta_{n,n\geq 1}(n+2\mu+1)\left(\frac{\left(\frac{1}{2}1+\frac{\kappa}{\alpha_s\beta_s}+2\mu\right)}{(\mu-1)}
      +\frac{1}{(\mu+1)}\right)
    \right]
  \right\}
}
}.
\end{equation}

\section{Radial probability density \label{sec4}}

Here, let us graphically analyze the behavior of the radial probability density $P(r)=P_{n,\kappa}(r)=\rho_{n,\kappa}(r)$ (or simply, probability density, which is the probability per unit radius or the probability of finding the DO at a radial distance $r$ from the origin) as a function of the radial coordinate/distance $r$ (i.e., $P(r)$ vs. $r$) for four different values of the quantum number $l$, angular frequency $\omega$, and of the curvature parameter $a$, with $n$ fixed, i.e., $n=0$ (ground state). In particular, to find $P(r)$, we must start from the curved total probability density (curved probability density per unit volume) given by $\bar{\Psi}_D \gamma^t (x)\Psi_D=\Psi^\dagger_D\gamma^0\gamma^t\Psi_D=\frac{1}{\alpha_s}\Psi^\dagger_D\Psi_D$. So, knowing that the curved volume element is given by $dV=\sqrt{-g(x)}d^3x=\beta^2_s r^2\sin\theta dr d\theta d\phi$, implies that the differential probability or total probability of finding the DO between $r$ and $r+dr$ is given by
\begin{equation}
dP_{total}=P(r)dr=\frac{1}{\alpha_s}\Psi^\dagger_D\Psi_D dV=\frac{\beta^2_s}{\alpha_s}\Psi_D^\dagger\Psi_D r^2\sin\theta dr d\theta d\phi.
\end{equation}

However, knowing that $\Psi_D=\frac{S\Phi}{r\sqrt{\sin\theta}}$, we have
\begin{equation}
P(r)dr=\frac{\beta^2_s}{\alpha_s}\Phi^\dagger \Phi dr d\theta d\phi,
\end{equation}
or better (using \eqref{spinor})
\begin{equation}
P(r)dr=\frac{\beta^2_s}{\alpha_s}\left\{[u(r)\chi^{m_j}_{\kappa}(\theta,\phi)]^2+[v(r)\chi^{m_j}_{-\kappa}(\theta,\phi)]^2\right\}dr d\theta d\phi.
\end{equation}

Now, integrating over all angles and knowing that the angular part of the spinor is already normalized, we have
\begin{equation}
P(r)dr=\frac{\beta^2_s}{\alpha_s}\int_0^{2\pi}\int_0^{\pi}\left\{[u(r)\chi^{m_j}_{\kappa}(\theta,\phi)]^2+[v(r)\chi^{m_j}_{-\kappa}(\theta,\phi)]^2\right\}dr d\theta d\phi=\frac{\beta^2_s}{\alpha_s}[u^2(r)+v^2(r)]dr.
\end{equation}
where implies
\begin{equation}
P(r)=\frac{\beta^2_s}{\alpha_s}[u^2(r)+v^2(r)].
\end{equation}

In other words, this is nothing more than the entire integrand of the integral \eqref{integral7}. Therefore, through the spinor \eqref{spinor5}, we have the following the radial components 
\begin{eqnarray}
&& u(r)=\bar{C}r^{2\mu+1/2}e^{-m_0\omega r^2/2}L^{2\mu}_{n}(m_0\omega r^2),
\\
&& v(r)=\frac{\bar{C}r^{2\mu-1/2}e^{-m_0\omega r^2/2}}{\left(E+\frac{m_0}{\alpha_s}\right)}\left[\left(\frac{1}{2}+\frac{\kappa}{\alpha_s\beta_s}+2\mu\right)L^{2\mu}_{n}(m_0\omega r^2)-2\delta_{n,n\geq 1}m_0\omega r^2 L^{2\mu+1}_{n-1}(m_0\omega r^2)\right],
\end{eqnarray}
where implies that $P(r)$ will be written as
\begin{equation}\label{P}
P(r)=\frac{\beta^2_s}{\alpha_s}\bar{C}^2 r^{4\mu-1}e^{-m_0\omega r^2}\left[r^2[L^{2\mu}_{n}(m_0\omega r^2)]^2+\frac{\left[\left(\frac{1}{2}+\frac{\kappa}{\alpha_s\beta_s}+2\mu\right)L^{2\mu}_{n}(m_0\omega r^2)-2\delta_{n,n\geq 1}m_0\omega r^2 L^{2\mu+1}_{n-1}(m_0\omega r^2)\right]^2}{\left(E+\frac{m_0}{\alpha_s}\right)^2}\right].
\end{equation}

However, for $n=0$ and $\kappa=l>0$, we have
\begin{equation}\label{P2}
P(r)=P_l(r)=\frac{\beta^2_s}{\alpha_s}\bar{C}^2 r^{4\mu-1}e^{-m_0\omega r^2}\left[r^2+\frac{\left(\frac{1}{2}+\frac{l}{\alpha_s\beta_s}+2\mu\right)^2}{\left(E+\frac{m_0}{\alpha_s}\right)^2}\right].
\end{equation}
or better (using \eqref{newnormalizationconstant})
\begin{equation}\label{P3}
P(r)=\
\frac{
  2(m_0\omega)^{2\mu+1}
}{
 (2\mu)!\left[
    1+\frac{m_0\omega}{\left(E+\frac{m_0}{\alpha_s}\right)^2}
    \frac{\left(\frac{1}{2}+\frac{l}{\alpha_s\beta_s}+2\mu\right)^2}{2\mu}
  \right]
}r^{4\mu-1}e^{-m_0\omega r^2}\left[r^2+\frac{\left(\frac{1}{2}+\frac{l}{\alpha_s\beta_s}+2\mu\right)^2}{\left(E+\frac{m_0}{\alpha_s}\right)^2}\right].
\end{equation}

Now, using $\mu$ (with $\kappa=l$) given in \eqref{newparameters}, and $E>0$ given in \eqref{spectra}, we obtain from \eqref{P3} the following probability densities for the original and modified metric
\begin{equation}\label{P4}
P(r)=\begin{cases}
P_{original}(r)=\frac{(\omega)^{\frac{l}{\sqrt{1-a}}+\frac{3}{2}}
}{\left(\frac{l}{\sqrt{1-a}}+\frac{1}{2}\right)!\left[\frac{1}{2}+\omega\frac{\left(\frac{2l}{\sqrt{1-a}}+1\right)}{\left(\sqrt{\frac{1}{(1-a)}+4\omega\left[\frac{l}{\sqrt{1-a}}+\frac{1}{2}\right]}+\frac{1}{\sqrt{1-a}}\right)^2}\right]}r^{\frac{2l}{\sqrt{1-a}}}e^{-\omega r^2}\left[r^2+\frac{\left(\frac{2l}{\sqrt{1-a}}+1\right)^2}{\left(\sqrt{\frac{1}{(1-a)}+4\omega\left[\frac{l}{\sqrt{1-a}}+\frac{1}{2}\right]}+\frac{1}{\sqrt{1-a}}\right)^2}
\right],
\\
P_{modified}(r)=\frac{(\omega)^{\frac{l}{\sqrt{1-a}}+\frac{3}{2}}
}{\left(\frac{l}{\sqrt{1-a}}+\frac{1}{2}\right)!\left[\frac{1}{2}+\omega\frac{\left(\frac{2l}{\sqrt{1-a}}+1\right)}{\left(\sqrt{1+4\omega\left[\frac{l}{\sqrt{1-a}}+\frac{1}{2}\right]}+1\right)^2}\right]}r^{\frac{2l}{\sqrt{1-a}}}e^{-\omega r^2}\left[r^2+\frac{\left(\frac{2l}{\sqrt{1-a}}+1\right)^2}{\left(\sqrt{1+4\omega\left[\frac{l}{\sqrt{1-a}}+\frac{1}{2}\right]}+1\right)^2}
\right].
\end{cases}
\end{equation}

\begin{figure}[!h]
\centering
\includegraphics[width=18.0cm]{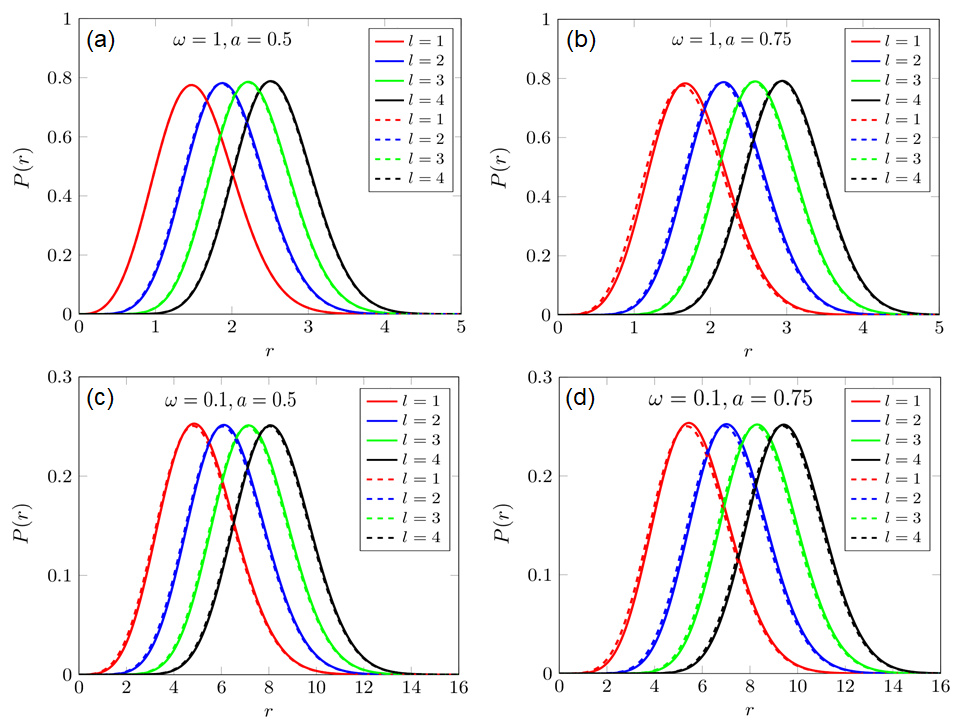}
\caption{Behavior of $P(r)$ vs. $r$ for four different values of $l$.}
\label{fig3}
\end{figure}

Therefore, in Fig. \ref{fig3}, we have the behavior of $P(r)$ vs. $r$ for four different values of $l$, where we use $\omega=1$ and $a=0.5$ (a), $\omega=1$ and $a=0.75$ (b), $\omega=0.1$ and $a=0.5$ (c), and $\omega=0.1$ and $a=0.75$ (d), being the solid curves for the original metric and the dashed curves for the modified metric. So, according to the figure, we see that the probability density for the two metrics is practically/approximately the same, that is, the (Gaussian-type) curves of the two metrics practically coincide ($P(r)_{original}\cong P_{modified}(r)$). In particular, in Figs. \ref{fig3}-(a) and \ref{fig3}-(b), the maximum values (height or peak values) of the curves increase very little with the increase of $l$, i.e., they have practically/approximately the same values. Already in Figs. \ref{fig3}-(c) and \ref{fig3}-(d), the maximum values (height or peak values) of the curves decrease very little with the increase of $l$, i.e., they also have practically/approximately the same values. Therefore, in both cases, we have $P_{l=1}(r)\cong P_{l=2}(r)\cong P_{l=3}(r)\cong P_{l=4}(r)$. Furthermore, the curves move to the right (or further and further away from the origin) as $l$ increases; that is, here, the main function of $l$ is to translate the probability density away from the origin (in other words, the higher the value of the orbital angular momentum, the greater the probability of finding the DO furthest from the origin). On the other hand, comparing (a) with (b), as well as (c) with (d), that is, the same angular frequency, but with different values of $a$ (or different curvatures), we see that the peak values are (slightly) higher and move away from the origin for a higher value of $a$ (or higher curvature). Now, comparing (a) with (c), as well as (b) with (d), that is, the same curvature, but with different values of $\omega$, we see that the peak values are higher and closer to the origin for a higher value of $\omega$ (in particular, this will become much clearer in the following figures/graphs).

\begin{figure}[!h]
\centering
\includegraphics[width=9.5cm]{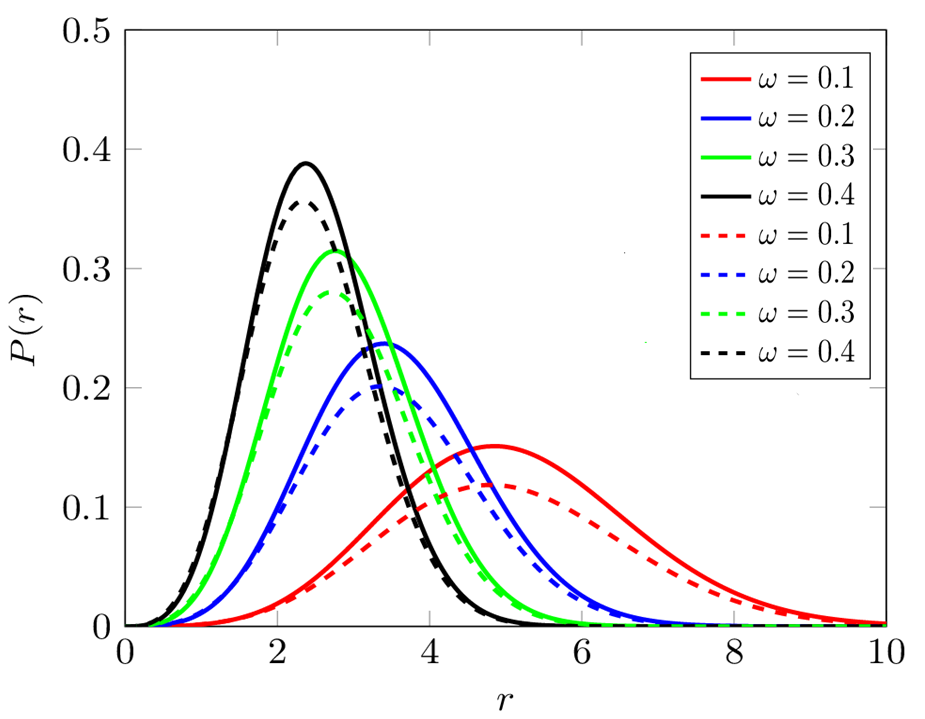}
\caption{Behavior of $P(r)$ vs. $r$ for four different values of $\omega$.}
\label{fig4}
\end{figure}

\begin{figure}[!h]
\centering
\includegraphics[width=9.5cm]{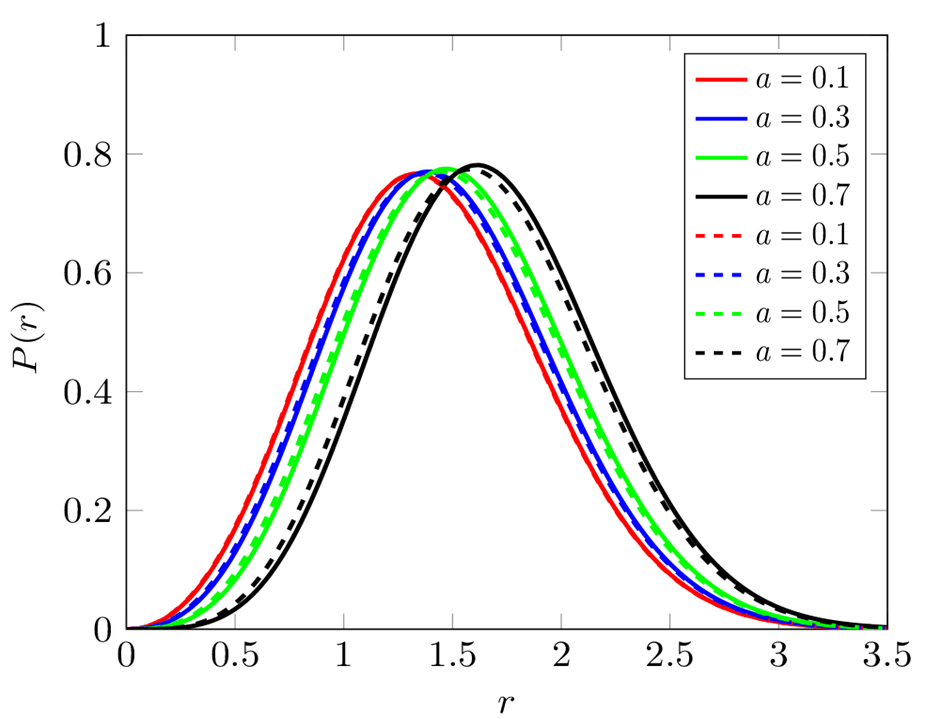}
\caption{Behavior of $P(r)$ vs. $r$ for four different values of $a$.}
\label{fig5}
\end{figure}

In Fig. \ref{fig4}, we have the behavior of $P(r)$ vs. $r$ for four different values of $\omega$, where we use $l=1$ and $a=0.5$, being the solid curves for the original metric and the dashed curves for the modified metric. So, according to the figure, we see that the probability density for the two metrics is not the same; that is, the maximum values (height or peak values) of the curves are higher for the original metric ($P_{original}(r)>P_{modified}(r)$). Besides, we see that the peak values are higher (and closer to the origin) for higher values of $\omega$. In other words, the greater the DO oscillation, the greater the probability density and the closer it will be to the origin. Already in Fig. \ref{fig5}, we have the behavior of $P(r)$ vs. $r$ for four different values of $a$, where we use $l=1$ and $\omega=1$, being the solid curves for the original metric and the dashed curves for the modified metric. So, according to the figure, we see that the probability density for the two metrics is practically/approximately the same, that is, the curves of the two metrics practically coincide ($P_{original}(r)\cong P_{modified}(r)$). Besides, we see that the peak values are (slightly) higher (and move away from the origin) for higher values of $a$. In other words, the greater the curvature of spacetime, the greater the probability density (but not that much) and the farther it will be from the origin (but not that much).


\section{Conclusions}\label{sec5}

In this paper, we determine the relativistic bound-state solutions for the DO in the curved spacetime (or under the gravitational effects) of a cloud of strings in $(3+1)$-dimensions, where such solutions are given by the four-component normalized Dirac spinor and by the relativistic energy spectrum (relativistic energy levels or high-energy spectrum). With the normalized spinor in hand, we were also able to determine another important result (often neglected in the literature), which is the radial probability density. However, unlike literature \cite{Saidi}, here, we work with two different forms/configurations for the spacetime/metric of a cloud of strings, that is, we work with the original form (developed by Letelier \cite{Letelier1,Letelier2,Letelier3}, where the parameter of the cloud of strings modifies both the temporal and radial parts of the line element/metric), as well as with the modified/rescaled form (the parameter modifies only the angular part of the line element/metric  \cite{Saidi,Costa}), respectively. To achieve our objective, we work with the curved DO in spherical coordinates, where the formalism used to write the DO in a curved spacetime was the tetrad formalism (or spin connection formalism/formulation) of RG (or gravity).

So, by defining a stationary ansatz for the spinor, we obtain two coupled first-order differential equations (which depend solely on the radial coordinate), and by substituting one equation into the other (i.e., decoupling one of the equations), we obtain a second-order differential equation (i.e., a Schrödinger-like equation or the ``decoupled/second-order radial DO''). To analytically and exactly solve this differential equation, we use a change of function (i.e., the original function of the equation was changed to a more suitable one), as well as a change of variable (i.e., the original variable of the equation was changed to a more suitable one). From this, we obtain the well-known Whittaker equation, whose solution is the Whittaker function, as well as a quantization condition. Consequently, we obtain the relativistic energy spectrum for the DO in the curved spacetime of a cloud of strings (in fact, we obtain two spectra, one for the original metric and the other for the modified metric), where such a spectrum describes/represents the positive-energy states/solutions (DO), as well as the negative-energy states/solutions (anti-DO), respectively. In particular, this spectrum has equal values (in module), i.e., both the DO and the anti-DO have positive energies as well as equal values, which implies that the spectrum is symmetric.

Besides, we note that the spectrum is quantized in terms of the radial quantum number $n$ and the angular quantum number $\kappa$ (or orbital quantum number $l$, since $\kappa=\mp (l\pm 1/2+1/2)$, with $\kappa_-=-l-1<0$ and $\kappa_+=l>0$), and explicitly depends on the angular frequency $\omega$ (describes the DO), curvature parameter $a$ or spacetime parameters $\alpha_s$ and $\beta_s$ (describes the cloud of strings), and on the effective rest mass $m_{\text{eff}}$ (describes the rest mass modified by curvature), respectively. About this effective rest mass, we see that it is only affected by the original metric; that is, the modified metric does not change the rest energy at all. However, the effect/influence of $\alpha_s$ and $\beta_s$ on the second term of the spectrum (or better, the second term within the square root of the spectrum) is equal, i.e., it does not matter if it is the original or modified metric; the second term will be the same. Consequently, this implies that the energies for the case of the original metric are greater than for the case of the modified metric ($E^{original}_{n,\kappa}>E^{modified}_{n,\kappa}$). So, analyzing the spectrum according to the values of $\kappa$, we note that for $\kappa>0$ (or spin down/$\downarrow$), the spectrum is quantized in terms of the quantum numbers $n$ and $l$ ($E^\downarrow_{n,\kappa>0}=E^\downarrow_{n,l}$), while for $\kappa<0$ (or spin up/$\uparrow$), it is quantized only in terms of $n$ ($E^\uparrow_{n,\kappa<0}=E^\uparrow_{n}$), respectively. Therefore, this implies that the energies for $\kappa>0$ are greater than for $\kappa<0$ ($\vert E_{n,l}^{\downarrow}\vert>\vert E_{n}^{\uparrow}\vert$). It is interesting to highlight that in the case of the modified metric for $\kappa<0$, the spectrum does not depend on the parameter $a$ (or $\beta_s$), i.e., it is as if the DO ``lived'' in the Minkowski spacetime. Also, the spectrum for $\kappa>0$ has its degeneracy broken (``undefined'' or not ``well-defined'') due to $\alpha_s$ and $\beta_s$ (``tied'' with $l$). It is also important to mention that in the absence of the cloud of strings, we recover exactly the usual spectrum of the DO in the $(3+1)$-dimensional Minkowski spacetime (as it should be).

Subsequently, we graphically analyze the behavior of the spectrum as a function of $\omega$ and $a$ for three different values of the quantum numbers $n$ and $l$ (since we chose $\kappa>0$). In particular, we performed this analysis in two different cases/scenarios, i.e., while $n$ varies ($n=0,1,2$), $l$ remains fixed ($l=1$), and while $l$ varies ($l=1,2,3$), $n$ remains fixed ($n = 0$), respectively. So, in graphs $E_n (\omega)$ vs. $\omega$ and $E_l(\omega)$ vs. $\omega$, we see that the behavior of the spectrum in both graphs is very similar, i.e., has the same profile/appearance (basically, the difference is only in the energy values, but except for $n=0$ and $l=1$, which appear in both graphs). Therefore, in both graphs, the energies increase with increasing $n$ and $l$ (as it should be), which implies that the energy difference (or the spacing) between two consecutive levels is positive. About this energy difference, it increases as a function of $\omega$, that is, the larger $\omega$, the greater the spacing between the energy levels. Besides, the energies (for a given
value of $n$ or $l$) also increase as a function of $\omega$, that is, $\omega$ has the objective of increasing the energies (as well as the
spacing between them). Consequently, this implies that the variation of energy as a function
of $\omega$ is positive.

Already in graphs $E_n (a)$ vs. $a$ and $E_l(a)$ vs. $a$, we see that the behavior of the spectrum in both graphs is also very similar, i.e., has the same profile/appearance (basically, the difference is only in the energy values, but except for $n=0$ and $l=1$, which appear in both graphs). Therefore, in both graphs, the energies increase with increasing $n$ and $l$ (as it should be); however, they tend to ``come together'' (or have the same value) for $a\to1$ (i.e., the greater the curvature of spacetime, the closer the energies will be to each other). In that way, the energy difference (or the spacing) between two consecutive levels is positive for $0\leq a<1$ and tends to zero for $a\to 1$, respectively. In particular, this energy difference is practically constant for $0\leq a \lesssim 0.5$, and decreases for $a>0.5$, in which it tends to zero for $a\to 1$. In other words, this is a consequence of the fact that the energies increase very little for $0\leq a \lesssim 0.5$ (the energies for the two metrics are practically the same) and a lot as $a$ increases (tending to infinity for $a\to 1$). Therefore, the energies increase as a function of $a$; however, this increase is only much more significant for increasingly larger values of $a$. Consequently, this implies that the variation of energy as a function of $a$ is practically zero for $0\leq a \lesssim 0.5$, and positive for $a>0.5$, in which it tends to infinity for $a\to 1$.

Lastly, we conclude our work through the graphical analysis of the behavior of the probability density $P(r)$ for four different values of $l$, $\omega$, and $a$, with $n=0$. So, for $l$ varying, we see that the probability density for the two metrics is practically the same, i.e., the curves of the two metrics practically coincide ($P(r)_{original}\cong P_{modified}(r)$). In particular, the graphs of $P(r)$ for $\omega=1$ and $0.5$ and for $\omega=1$ and $0.75$, the maximum values of the curves increase very little with the increase of $l$, i.e., they have practically the same values. Already in the graphs of $P(r)$ for $\omega=0.1$ and $0.5$ and for $\omega=0.1$ and $0.75$, the maximum values of the curves decrease very little with the increase of $l$, i.e., they also have practically the same values. Also, the curves move to the right (or further and further away from the origin) as $l$ increases, i.e., the main function of $l$ is to translate the probability density away from the origin. Now, for $\omega$ varying, we see that the probability density for the two metrics is not the same, i.e., the maximum values of the curves are higher for the original metric ($P_{original}(r)>P_{modified}(r)$). Also, we see that the peak values are higher (and closer to the origin) for higher values of $\omega$. In other words, the greater the DO oscillation, the greater the probability density and the closer it will be to the origin. Finally, for $a$ varying, we see that the probability density for the two metrics is practically the same, i.e., the curves of the two metrics practically coincide ($P_{original}(r)\cong P_{modified}(r)$). Also, we see that the peak values are higher (and move away from the origin) for higher values of $a$. In other words, the greater the curvature of spacetime, the greater the probability density and the farther it will be from the origin.

\section*{Acknowledgments}

\hspace{0.5cm}

The author would like to thank the anonymous referee for their careful reading of the article, which contributed substantially to improving its quality, as well as the Conselho Nacional de Desenvolvimento Cient\'{\i}fico e Tecnol\'{o}gico (CNPq) for financial support.

\section*{Data availability statement}

\hspace{0.5cm} This manuscript has no associated data or the data will not be deposited. [Authors’ comment: Data sharing not applicable to this article as no datasets were generated or analysed during the current study].

\section*{Code Availability Statement}

\hspace{0.5cm} This manuscript has no associated code/software. [Author’s comment: Code/Software sharing not applicable to this article as no code/software was generated or analysed during the current study].

\end{document}